\documentclass[12pt]{article}
\usepackage{multirow}
\usepackage{threeparttable}
\usepackage[table,xcdraw]{xcolor}
\usepackage{dcolumn} 
\usepackage{booktabs}
\usepackage{float}

\usepackage{amssymb, amsthm, amsmath}
\usepackage{dsfont}
\usepackage{bigints}
\usepackage{nameref}

\usepackage{graphicx}
\usepackage{subcaption}

\usepackage{comment}

\newcommand{\argmax}{\operatornamewithlimits{argmax}}
\newcommand{\argmin}{\operatornamewithlimits{argmin}}

\RequirePackage[colorlinks,citecolor=blue,linkcolor=blue,urlcolor=blue]{hyperref}

\usepackage[backend=bibtex,style=authoryear]{biblatex}
\theoremstyle{plain}

\newtheorem{assumption}{Assumption}

\theoremstyle{remark}
\newtheorem{definition}{Definition}[section]

\begin{document}

\begin{titlepage}
\title{Strategic Waiting in Centralized Matching: Daycare Assignment}
\author{Kan Kuno \thanks{kankuno@e.u-tokyo.ac.jp. Acknowledgments: I extend my sincere gratitude to my PhD advisor, Karl Schurter, Paul Grieco, and Ran Shorrer, for their invaluable guidance. I am grateful to Fuhito Kojima, Marc Henry, and Kala Krishna for their insightful feedback. I also appreciate the contributions from the participants of Penn State’s Applied Micro Brown Bag, SWET IO 2024, and APIOC 2024. Special appreciation goes to the Bunkyo municipality of Tokyo for their crucial data support.}}
\date{\today}
\maketitle
\begin{abstract}
\noindent In centralized assignment mechanisms, unassigned participants are often placed on waitlists to improve their chances in future rounds. However, I demonstrate that this practice may bring unintended welfare consequences on the participants, using data from the Japanese daycare system. The prioritization of waitlisted applicants introduces a dynamic incentive for applicants to manipulate their priority by strategically choosing to be waitlisted to secure positions at more selective daycares. I show that 30.0\% of applicants do not list safety options when they initially apply and that those who benefit from additional priority are 8.1 percentage points more likely to avoid listing safety options than those who do not. Given the prevalence of such strategic waiting, I estimate a structural model of daycare choice that extends Agarwal and Somaini (2018) to a two-period model allowing for reapplication. By simulating a scenario without waitlist priority, I find that the current priority functions as a redistributive mechanism: through its abolition, early starters (age 0) experience a 34.1\% decrease in welfare and a 1.7 percentage point increase in the likelihood of being waitlisted, whereas late starters (age 1) experience a 29.0\% increase in welfare and a 6.9 percentage point decrease in the likelihood of being waitlisted.

\vspace{0in}

\bigskip
\end{abstract}

\setcounter{page}{0}
\thispagestyle{empty}
\end{titlepage}
\pagebreak \newpage

\section{Introduction} \label{sec:introduction}

A centralized assignment mechanism cannot always ensure that all participants are assigned. As a remedial measure, unassigned participants are typically placed on a waitlist to improve their chances in subsequent rounds. While this approach appears equitable, what are its welfare implications? Using data from the Japanese centralized daycare assignment system, which shares similarities with the U.S. school choice system, I demonstrate that this practice may lead to mixed or unintended outcomes for participants. The core issue lies in the prioritization of waitlisted applicants, which creates a dynamic incentive for applicants to manipulate their own priority by intentionally remaining unassigned to increase their chances of securing a spot at their preferred, often more selective, daycare.

When matching opportunities arise dynamically over time, agents may have incentives to strategically delay their decisions in hopes of securing a better match. This behavior has been documented in various areas of market design, including organ transplants (\cite{agarwalEquilibriumAllocationsAlternative}), public housing (\cite{waldingerTargetingInKindTransfers2021a}), and license allocation (\cite{verdierWelfareEffectsDynamic2022}). In the context of school choice, students are assigned to schools based on their priorities and reported preferences through a specific algorithm. While existing studies have primarily focused on inferring students' true preferences from their reported ones, these analyses often treat priorities as fixed. However, when the mechanism prioritizes waitlisted participants, it introduces incentives for strategic waiting: applicants can manipulate their probability of being waitlisted and their future priority through their reported preferences. This dynamic not only complicates the interpretation of preferences but also challenges the welfare assessment of the mechanism.

The welfare consequences of strategic waiting are of significant policy interest in this context. The Japanese daycare market has been grappling with a severe shortage of seats in daycare centers, sparking widespread public concern and heated political debate. The issue gained national attention in early 2016, when a viral blog post titled "\textit{Hoikuen Ochita Nihon Shine!!!}" (My child wasn’t accepted for nursery school. Die, Japan!!!) resonated with frustrated parents across the country.\footnote{Mainichi Shimbun, "Parents protest over lack of child care facilities after Abe's weak response to blog post," March 8, 2016. Available at: \url{https://mainichi.jp/english/articles/20160308/p2a/00m/0na/015000c}} Competition for available seats remains intense: in Tokyo's Bunkyo municipality, more than one in four applicants who applied to age 0 classes were unassigned and waitlisted in 2019. To address this, waitlisted applicants are granted additional points to their priority scores when they reapply.
While this may at first sight seem to decrease the number of unassigned applicants, anecdotal evidence suggests that it has brought unintended, if not opposite, consequences. 
There have been several blog posts advising average-score applicants to earn these extra points by choosing to be on the waitlist if they wish to be admitted to more selective centers. 
For example, Kanako Mishima, a working mother of two children, shares her experience in her blog post:\footnote{Source: FP Mishima, \url{https://fpmishima.com/2020/03/05/hokatsuhtml}. Accessed July 4, 2024. Translated using Google Translate.}
\begin{quote}
    There are two factors that you can control to raise your adjustment score: use of non-accredited daycare center and being waitlisted. You would typically need 26-28 points to enter... Since these cutoffs do not change dramatically every year, ...you should do whatever you can do beforehand... I also put my two children into a non-accredited center... and was able to put both of them into my first choice!
\end{quote}
Is such strategic waiting a practical issue or just a theoretical possibility? How does the waitlist priority affect overall competition and welfare? The answers to these questions are not ex-ante obvious, and are empirical questions.

This paper first documents the prevalence of strategic waiting driven by waitlist prioritization, using application and assignment data from Tokyo's Bunkyo municipality. Through a numerical example, I argue that the key predicted pattern in reported preferences is that applicants strategically avoid listing safety options during their initial application, aiming to increase their chances of admission to more selective daycare centers in the next period. The data supports this prediction: 30.0\% of applicants do not list safety options when they first apply. To rule out alternative explanations for this pattern, such as the declining value of the outside option as the child ages, I analyze how the benefits of extra priority vary among applicants. For example, an applicant with a high initial priority score has little incentive to risk being unassigned for additional priority, as their score is already sufficient to gain admission to their preferred daycare. Similarly, applicants with very low initial priority scores are unlikely to avoid listing safety options, since even with extra priority, their chances of admission to selective daycares remain low. It is the applicants for whom extra priority is both necessary and sufficient to secure admission to their preferred daycare who are most likely to strategically avoid safety options. To test this effect, I develop a measure capturing how extra priority increases the likelihood of admission to selective daycares. Using this measure, I find that applicants who benefit from being waitlisted are 25.7\% more likely to avoid listing safety options than others.

Motivated by the significance of these dynamic incentives, I estimate a structural model of daycare choice to recover applicants' underlying preferences. Building on \cite{agarwalDemandAnalysisUsing2018}'s static framework of school choice, I extend the model to account for the possibility that an applicant may reapply the following year if waitlisted in the initial round of applications. 
Assuming that preferences remain stable over time, the problem simplifies to a static choice framework, where applicants select a single lottery that determines their assignment probabilities across daycares based on their initially and subsequently reported preferences. The estimation process involves two stages. In the first stage, I estimate the lotteries induced by each possible pair of rank-ordered lists (ROLs) using the bootstrap estimator developed by \cite{agarwalDemandAnalysisUsing2018}. In the second stage, I estimate the preference parameters through the method of simulated moments. 
Given the high cardinality of the choice set, I propose a heuristic algorithm to approximate the optimal pair of ROLs for a given preference profile.

Given my structural estimates I simulate a counterfactual scenario in which no additional priority is granted for being waitlisted.
The counterfactual experiments highlight the redistributive effects of waitlist prioritization in daycare admissions, showing how changes in priority impact cutoff distributions, welfare outcomes, and application behaviors. For age 0 applicants, reduced waitlist priority leads to more daycare centers having cutoffs rather than being under-enrolled, but with little shift in their distribution's peak. Conversely, for age 1 and age 2 applicants, reduced priority shifts cutoff distributions significantly, lowering the barriers to admission. Welfare analysis reveals that waitlist priority benefits applicants who arrive early to the market at age 0, especially those with lower initial scores, by increasing their chances of admission in subsequent periods. However, it disadvantages applicants who arrive lately at age 1, whose utility declines due to heightened competition from waitlisted cohorts. These findings demonstrate that waitlist priority, while designed to support unassigned applicants, acts as a redistributive mechanism, favoring early starters at the expense of late starters.

My results also provide insights into school choice systems more broadly. While the prioritization of waitlisted re-applicants over new applicants might be unique to Japanese daycare, some U.S. school choice systems also use waitlisting, creating similar, if not identical, incentives for strategic waiting among participants.
For example, an article on Delaware’s school choice system describes how their waitlist process “creates a waiting-game, where parents are holding out to hear back from their other choices, which might have been a priority over the school they did get accepted to.”\footnote{Delaware Live, "What happens if you’re waitlisted from school of choice?" Available at: \url{https://delawarelive.com/what-happens-if-youre-waitlisted-school-of-choice/}}
The opportunity to secure a top-choice school by remaining on the waitlist encourages parents to delay accepting offers from less-preferred schools. This mirrors the mechanism studied in this paper, where applicants strategically adjust their behavior to take advantage of waitlist prioritization.

\section{Literature} \label{sec:literature_review}

This paper contributes to several areas of literature. First, it extends empirical studies on school choice and college admission mechanisms, which have largely focused on analyzing the distribution of underlying preferences. Unlike approaches that assume reported preferences reflect true preferences, \cite{fackTruthTellingPreferenceEstimation2019} estimates preferences using the mechanism's stability property. In contrast, \cite{agarwalDemandAnalysisUsing2018}, \cite{calsamigliaStructuralEstimationModel2020}, and \cite{kaporHeterogeneousBeliefsSchool2020} consider the possibility that reported preferences may be manipulated, rather than assuming truth-telling or stability. Building on \cite{agarwalDemandAnalysisUsing2018}, who introduces a two-step estimation process—first estimating assignment probabilities and then preference parameters—this paper makes a further contribution by addressing a different reason why reported preferences should not be assumed to be truthful: they may be strategically selected to influence one’s priorities. By showing that student priority can be affected through the manipulation of reported preferences, this paper proposes an estimation framework that relaxes the assumption of fixed priority and accounts for this strategic behavior.

To the best of my knowledge, the only empirical studies on school choice or college admissions that incorporate dynamic elements are \cite{naritaMatchMismatchLearning2018}, \cite{larroucauDynamicCollegeAdmissions}, and \cite{hahmDynamicFrameworkSchool}. \cite{naritaMatchMismatchLearning2018} examines New York City's school choice system, quantifying student learning and inertia under the assumption of truth-telling. Similarly, \cite{hahmDynamicFrameworkSchool} analyzes the impact of middle school choice on high school choice within the same system, assuming stability. My paper, however, addresses a context where applicants may omit listing their safety options from their reported preferences, making the assumptions of truth-telling or stability problematic. \cite{larroucauDynamicCollegeAdmissions}, on the other hand, focuses on Chilean college admissions, developing a dynamic model of school choice that allows students to update their preferences based on their performance, modeling the application decision as an optimal portfolio problem. Unlike their focus on initial mismatches and student learning, this paper emphasizes the dynamic incentives generated by the mechanism itself.

This paper also contributes to empirical studies of dynamic allocation systems and waitlist design, with organ transplants and public housing being the primary focus areas of previous literature. \cite{agarwalEquilibriumAllocationsAlternative} document that patients strategically wait for better organs, modeling this as an optimal stopping problem. \cite{sweatEndogenousPriorityCentralized} examines the endogeneity of priority in the U.S. heart transplant waitlist, finding that a policy change improved survival for sicker patients but led to greater selectivity in transplant offers, ultimately redistributing rather than increasing aggregate survival. In public housing, \cite{waldingerTargetingInKindTransfers2021a} shows how waiting time functions as a price, balancing the trade-off between choice and targeting, where eliminating choice improves targeting but reduces tenant welfare. Similarly, \cite{leeDynamicAllocationPublica} demonstrates that increasing public housing supply doesn't reduce wait times because households strategically delay applying, and proposes a strategyproof mechanism to address this. My work contributes to this literature by being the first to investigate the welfare impacts of waitlist priority, which could further incentivize strategic waiting. This adds to the broader discussion on the design of effective waitlist systems.

Finally, this paper contributes to the study of matching in daycare markets, marking the first empirical application in this area. \footnote{Beyond market design, studies on Japanese childcare, such as \cite{asaiChildcareAvailabilityHousehold2015}, \cite{yamaguchiEffectsSubsidizedChildcare2018}, \cite{yamaguchiHowDoesEarly2018}, \cite{yamaguchiEffectsParentalLeave2019}, and \cite{fukaiChildcareAvailabilityFertility2017}, focus on impacts on maternal employment, child development, and fertility.} Empirical work on daycare market matching is scarce, with most contributions being theoretical. \cite{kennesDayCareAssignment2014} examine the centralized allocation of children to public daycare centers in Denmark, addressing dynamic matching with entry and exit over time. They demonstrate that no mechanism is both stable and strategy-proof, proposing instead a strategy-proof and Pareto-efficient mechanism where parents sequentially choose menus of schools. \cite{kamadaFairMatchingConstraints2023} prove the existence of a student-optimal matching that is feasible, individually rational, and fair for a class of school choice problems. They simulate their matching concept using rank-order list (ROL) data on daycare assignments from the same municipality as the data used in this project, assuming truth-telling by students.

\section{Institutional Details} \label{sec:institutional_details}
The Japanese daycare market is a regulated and subsidized market, where instead of a flexible price mechanism like the US, users are allocated to each center by a centralized matching mechanism. Children between ages 0 and 5 can attend a daycare center, before starting primary education. There are accredited and non-accredited daycare centers, and most users of daycare services use accredited centers. Each municipality runs its own matching mechanism to allocate applicants to accredited centers.

This paper examines data from Bunkyo-ku, a municipality in central Tokyo with a population of over 200,000. The daycare application process operates as follows: To secure a daycare spot starting in April, the beginning of the academic year, applicants must apply to the municipality in November. Prior to the application period, each daycare center reports the number of available seats for five age groups (0, 1, 2, 3, and 4–5), and these figures are made publicly available on the municipality’s website. Based on this information, applicants submit a ranked order list (ROL) of their top five preferred daycare centers. Each applicant is assigned a priority score, determined by factors such as the parents’ employment status and other relevant criteria. An algorithm matches applicants to daycare centers based on their ROLs and priority scores for each age group. Applicants who are not matched to any center are placed on a waiting list. The system allows for reapplications, giving waitlisted applicants the opportunity to reapply in future applications. Current users of accredited daycare centers are not involuntarily displaced but may apply to transfer to a different center if they wish.

The assignment algorithm used is known as the (truncated) serial dictatorship algorithm, details of which are shown below:
\begin{description}
    \item[] Step 0: List applicants according to their priority scores, from the highest to the lowest. Ties are broken deterministically according to another priority rule\footnote{See Appendix \ref{tab:priorityscore}.}.
    \item[] Step $k \geq 1$: For the student $k$th highest in the list, assign her to the center which, among the centers with at least one vacant seat still left, is ranked the highest in her ROL. The vacant seat of the center she is assigned to decreases by one. If she is not assigned to any center, she is waitlisted.
\end{description}
The application guide provided by the municipality emphasizes that applicants are neither prioritized nor punished based on where they rank a particular center, the number of centers they list, or the timing of their application, as long as it is submitted before the deadline.

The priority score of an applicant is calculated as the sum of three components: the mother's basic score, the father's basic score, and an adjustment score. The basic score primarily reflects the working status of each parent, while the adjustment score accounts for other factors that influence the need for childcare services. A typical applicant might have both parents working (+10 each), reside within the municipality (+4), not currently use an accredited daycare center (+1), and lack nearby grandparents who could provide childcare (+1), resulting in a total score of 26.
Applicants can easily calculate their own priority scores using a table provided in the application guide, also shown in Appendix \ref{tab:priorityscore}. Notably, applicants who are unassigned in their first application can increase their priority score by two points when they reapply: one point for being on the waitlist and another point for using a non-accredited center while waiting. Conversely, applicants already assigned to a daycare center lose the additional point granted to first-time applicants if they reapply.
Although detailed data on the components of each applicant's priority score is unavailable, aggregate data from the municipality indicates that 92.16\%, 97.93\%, and 99.75\% of waitlisted applicants in 2019, 2020, and 2021, respectively, utilized non-accredited centers while on the waitlist. Since being placed on the waitlist occurs whenever an applicant is unassigned, it is reasonable to assume that such applicants would have had a two-point lower priority score had they not been waitlisted, assuming all other factors affecting their scores remained constant.

\section{An Illustrative Example}
\label{sec:illustrative_example}
In this section, using a simple numerical example I will demonstrate the key observable pattern of rank-ordered lists (ROLs) that can arise from applicants strategically becoming waitlisted, as well as its relation to alternative potential reasons driving such pattern. 

Suppose there are two periods, \(t = 1, 2\), and two daycare centers \(A\) and \(B\). An applicant receives a constant flow utility \(v_j\) from attending center \(j\). Center \(A\) is more popular, providing a higher flow utility than center \(B\): \(v_A = 7 > v_B = 2\). In period 1, the applicant chooses a ROL \(R^1 \in \{(AB), (A), (BA), (B)\}\). Here, \((AB)\) is a ROL where \(A\) is ranked first and \(B\) second, \((A)\) lists only \(A\), and \((BA)\) and \((B)\) are defined similarly. The applicant faces randomness in admission chances: let \(p^1_j\) denote the probability of being admitted to center \(j\) if \(j\) is the only center listed in \(R^1\). Center \(A\), being more popular, is more selective: \(p^1_A = 0.1 < p^1_B = 0.5\). The admission chances are assumed to be independent across centers. This implies that if the applicant chooses \(R^1 = (AB)\), she will be assigned to center \(A\) with probability \(p^1_A\) and to center \(B\) with probability \((1 - p^1_A) p^1_B\). If the applicant is not admitted to either center, she will be waitlisted. A waitlisted applicant can reapply in period 2, choosing \(R^2 \in \{(AB), (A), (BA), (B)\}\). When waitlisted, the applicant's admission chances improve: \(p^2_A = 0.5\) and \(p^2_B = 0.9\). In both periods, if the applicant is unassigned, she uses the outside option, which has a flow utility normalized to zero. Panel A of Table \ref{tab:num_illus} summarizes the flow utilities and individual admission chances in the two periods.


\begin{table}[htbp!]
\centering

\begin{threeparttable}
\begin{tabular}{c}
{Panel A: Preferences and Admission Chances} \\
\end{tabular}

\begin{tabular}{cccc}
\hline
Center  & $v_j$ & $p^1_j$  & $p_j^2$ \\ \hline
A & 7  & 0.1 & 0.5 \\
B & 2  & 0.5 & 0.9 \\
\hline
\end{tabular}

\begin{tabular}{c}
{Panel B: Assignment and Waitlist Probabilities, and Expected Flow Utilities} \\
\end{tabular}

\begin{tabular}{lcccc}
\hline
 & (AB) & (A) & (BA) & (B) \\
\hline
\multicolumn{5}{l}{{Period t=1}} \\
Assignment probability to center A & $ 0.10 $ & $ 0.10 $ & $ 0.05 $ & $ 0.00 $ \\
Assignment probability to center B & $ 0.45 $ & $ 0.00 $ & $ 0.50 $ & $ 0.50 $ \\
Waitlist probability & $ 0.45 $ & $ 0.90 $ & $ 0.45 $ & $ 0.50 $ \\
Expected flow utility & $ 1.60 $ & $ 0.70 $ & $ 1.35 $ & $ 1.00 $ \\
\hline
\multicolumn{5}{l}{{Period t=2}} \\
Assignment probability to center A & $ 0.50 $ & $ 0.50 $ & $ 0.05 $ & $ 0.00 $ \\
Assignment probability to center B & $ 0.45 $ & $ 0.00 $ & $ 0.50 $ & $ 0.90 $ \\
Waitlist probability & $ 0.05 $ & $ 0.50 $ & $ 0.45 $ & $ 0.10 $ \\
Expected flow utility & $ 4.40 $ & $ 3.50 $ & $ 1.35 $ & $ 1.80 $ \\
\hline
\end{tabular}

\begin{tablenotes}
    \scriptsize
    \item \textit{Note:} 
    The table illustrates a numerical example with two periods (\(t = 1, 2\)) and two daycare centers (\(A\) and \(B\)).
    Panel A summarizes the flow utilities (\(v_j\)) and admission probabilities (\(p_j^t\)) for each center. \(v_j\) represents the utility derived from attending center \(j\). \(p_j^1\) and \(p_j^2\) denote the probabilities of being admitted to center \(j\) in periods 1 and 2, respectively. 
    Panel B presents the assignment probabilities, waitlist probabilities, and expected flow utilities associated with different rank-ordered lists (ROLs) in each period. The ROLs include \((AB)\) (center \(A\) ranked first, \(B\) second), \((A)\) (only \(A\) listed), \((BA)\) (center \(B\) ranked first, \(A\) second), and \((B)\) (only \(B\) listed). The expected flow utility is computed as the weighted average of \(v_A\) and \(v_B\), based on the assignment probabilities for each ROL in a given period.
\end{tablenotes}
    
\end{threeparttable}

\caption{Numerical Example}
\label{tab:num_illus}
\end{table}

Given this setup, the applicant's expected flow utility from a given ROL in a given period can be computed as the weighted mean of \(v_A\) and \(v_B\), with weights equal to the assignment probability implied by the ROL in that period. For example, the applicant's expected flow utility in period 1 from submitting \(R^1 = (AB)\) is \(p^1_A v_A + (1 - p^1_A) p^1_B v_B\). Panel B of Table \ref{tab:num_illus} summarizes the assignment probability to each center and the expected utility associated with each possible ROL in each period. If the applicant myopically chooses \(R^1\) and \(R^2\) to maximize the expected utility in each period, she would choose \(AB\) in both periods: she simply lists her most favored choice \(A\) first and also lists her {safety option} \(B\) second. On the other hand, suppose the applicant is forward-looking and chooses \(R^1\) and \(R^2\) to maximize the discounted sum of expected utility in both periods, discounting period 2 utility by a factor of \(\delta = 0.99\). While she would still choose \(R^2 = (AB)\), she would now choose \(R^1 = (A)\) instead of \((AB)\). This is because, by dropping her safety option \(B\) from \(R^1\), she can increase her probability of being waitlisted in period 1 from \(0.45\) to \(0.90\). By being waitlisted, she can reapply in period 2, under which she faces a much higher admission chance to her favorite center \(A\). This benefit more than compensates for the loss in expected flow utility in period 1. This pattern of applicants dropping one's safety option only in her initial round of application ($R^1 = (A)$ then $R^2 = (AB)$) is observable if I have data on individual ROLs and priority scores.

One might argue that other factors could explain the pattern of dropping safety options. For instance, it could be suggested that an applicant's outside option worsens between period 1 and period 2, or that the applicant was initially overconfident about her admission chances and only applied to a safety option after adjusting her beliefs following a rejection in period 1. Both scenarios could result in a similar truncation pattern of ROLs, even if the applicant were myopic. However, in the numerical example above, the applicant modified her ROL solely because the increased waitlist priority improved her chances of being admitted to her preferred daycare: \(p^2_1 > p^1_1\). Without this increase in admission chances, she would have continued to submit \(R^1 = (AB)\). The likelihood that waitlist priority boosts an applicant's chances of admission to popular daycares is unlikely to be correlated with either the probability of outside options deteriorating with age or the applicant’s level of naivety regarding cutoffs. Therefore, if this improvement in admission chances is positively associated with the observed pattern of dropping safety options, it strongly suggests that applicants are responding to strategic incentives to become waitlisted.

\section{Data} \label{sec:data}

I use applicant-level data from Bunkyo Municipality's daycare division, covering the years 2019 to 2021. The dataset includes key variables such as the age of the applicant's child (ranging from 0 to 5), priority score, rank-ordered list (ROL), and one of eight area codes (A to H) corresponding to the applicant's residence. Area H represents applicants residing outside the municipality, comprising less than 5.79\% of the sample. Additionally, the dataset records the daycare center to which each applicant was assigned. Applicants who reapply in multiple years are given unique identifiers, enabling the tracking of reapplications. From the municipality's website, I obtain which includes information for each center-year-age combination: the number of vacant seats, the cutoff score, and the total number of applications. 

Although not directly utilized in the structural estimation, I incorporate data on daycare center characteristics from various sources. First, I collect information from the municipality, including exact addresses, total capacity, whether the center is publicly operated, and whether it qualifies as a certified center that can also serve educational purposes\footnote{A daycare qualified as such is called. \textit{Centers for early childhood education and care}, which refers to centers combining daycare and educational functions, recognized under Japanese law.}.
Next, I gather data on each daycare's nearest train station and its walking distance using \url{https://www.benricho.org/}, reflecting the popularity of train commuting in central Tokyo. Additionally, I obtain information on the total number of full-time staff, whether the center has undergone a third-party review,\footnote{This is a system where an external organization assesses the operational status and quality of services provided by daycare centers. The process involves collecting feedback from both staff and parents, conducting on-site inspections, and publishing the results to ensure transparency and foster trust in childcare services.} and whether it offers temporary childcare services\footnote{This is a program designed to provide short-term care for infants and young children when their families face temporary difficulties in caregiving. This service is primarily offered during daytime hours at daycare centers.} from the database of daycare centers provided by \url{https://www.wam.go.jp/kokodesearch/}.
These variables are employed to validate the plausibility of daycare fixed effects in estimating applicants' preferences.

\begin{figure}[htbp]
    \centering
    \includegraphics[width = 0.75 \textwidth]{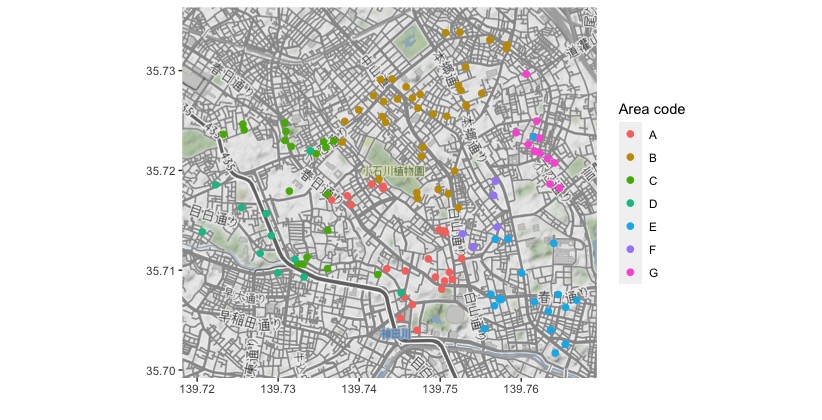}
    \caption{Location of Accredited Daycare Centers}
    \medskip 
    \begin{minipage}{0.75\textwidth} 
        {\scriptsize \textit{Note:} Alphabets A to H correspond to the area codes of each daycare center in the data.
        \par}
    \end{minipage}
    \label{fig:map}
\end{figure}

In total, the dataset covers 126 daycare centers and 6,450 applications. Figure \ref{fig:map} provides a map showing the locations of the daycare centers by area code. Descriptive statistics for daycare centers and applicants in 2020 are presented in Tables \ref{tab:sumstat_schools_2020} and \ref{tab:sumstat_students_2020}, with those for other years available in Tables \ref{sumstat_schools_full} and \ref{sumstat_students_full}. In the following, I explain some key aspects of the data that motivate my modelling approach.

\begin{table}[htbp!]
\centering
\caption{Summary Statistics for Daycare Centers (Year 2020)}
\begin{tabular}{lcccccc}
  \hline
 & Age 0 & Age 1 & Age 2 & Age 3 & Age 4 & Age 5 \\ 
  \hline
Number of centers & 95 & 112 & 112 & 86 & 77 & 77 \\ 
  Number of public centers & 16 & 21 & 21 & 20 & 17 & 17 \\ 
  Total capacity & 661 & 1137 & 1272 & 1263 & 2404 & 2404 \\ 
  Mean capacity & 6.96 & 10.15 & 11.36 & 14.69 & 31.22 & 31.22 \\ 
  Number of centers with vacant seats & 95 & 100 & 80 & 55 & 49 & 29 \\ 
  Total vacant seats & 645 & 574 & 286 & 284 & 293 & 181 \\ 
  Mean vacant seats & 6.81 & 5.75 & 3.57 & 5.24 & 6.10 & 6.24 \\ 
  Mean total applications & 28.42 & 32.25 & 12.46 & 17.71 & 6.22 & 2.00 \\ 
  Mean top ranked applications & 7.30 & 7.73 & 3.14 & 4.30 & 1.35 & 0.38 \\ 
  Fraction of no cutoffs & 0.33 & 0.03 & 0.33 & 0.36 & 0.71 & 0.82 \\ 
  Mean cutoff score & 25.74 & 26.43 & 26.81 & 27.34 & 27.50 & 26.80 \\ 
   \hline
\end{tabular}
\label{tab:sumstat_schools_2020}
\end{table}

\begin{table}[htbp!]
\centering
\caption{Summary Statistics for Applicants (Year 2020)}
\begin{tabular}{lcccccc}
  \hline
 & Age 0 & Age 1 & Age 2 & Age 3 & Age 4 & Age 5 \\ 
  \hline
Number of applications & 660 & 774 & 275 & 290 & 122 & 40 \\ 
  Mean priority score & 26.45 & 26.64 & 26.36 & 26.95 & 27.34 & 27.25 \\ 
  Fraction of already waitlisted applicants & 0.000 & 0.169 & 0.342 & 0.121 & 0.115 & 0.300 \\ 
  Fraction of incumbent applicants & 0.002 & 0.047 & 0.196 & 0.217 & 0.246 & 0.200 \\ 
  Mean list length & 4.00 & 4.07 & 3.80 & 3.83 & 3.08 & 2.65 \\ 
  Fraction of top choice in the same area & 0.269 & 0.277 & 0.228 & 0.144 & 0.176 & 0.000 \\ 
  Fraction of being assigned to first choice & 0.582 & 0.416 & 0.385 & 0.390 & 0.533 & 0.350 \\ 
  Fraction of being assigned to second choice & 0.158 & 0.146 & 0.229 & 0.224 & 0.098 & 0.050 \\ 
  Fraction of being assigned to third choice & 0.068 & 0.083 & 0.076 & 0.076 & 0.041 & 0.075 \\ 
  Fraction of being assigned to fourth choice & 0.038 & 0.053 & 0.044 & 0.045 & 0.033 & 0.050 \\ 
  Fraction of being assigned to fifth choice & 0.012 & 0.031 & 0.022 & 0.010 & 0.008 & 0.025 \\ 
  Fraction of ending up unassigned & 0.142 & 0.247 & 0.171 & 0.110 & 0.107 & 0.350 \\ 
   \hline
\end{tabular}
\label{tab:sumstat_students_2020}
\end{table}

\begin{figure}
    \centering
    \includegraphics[width=0.9\linewidth]{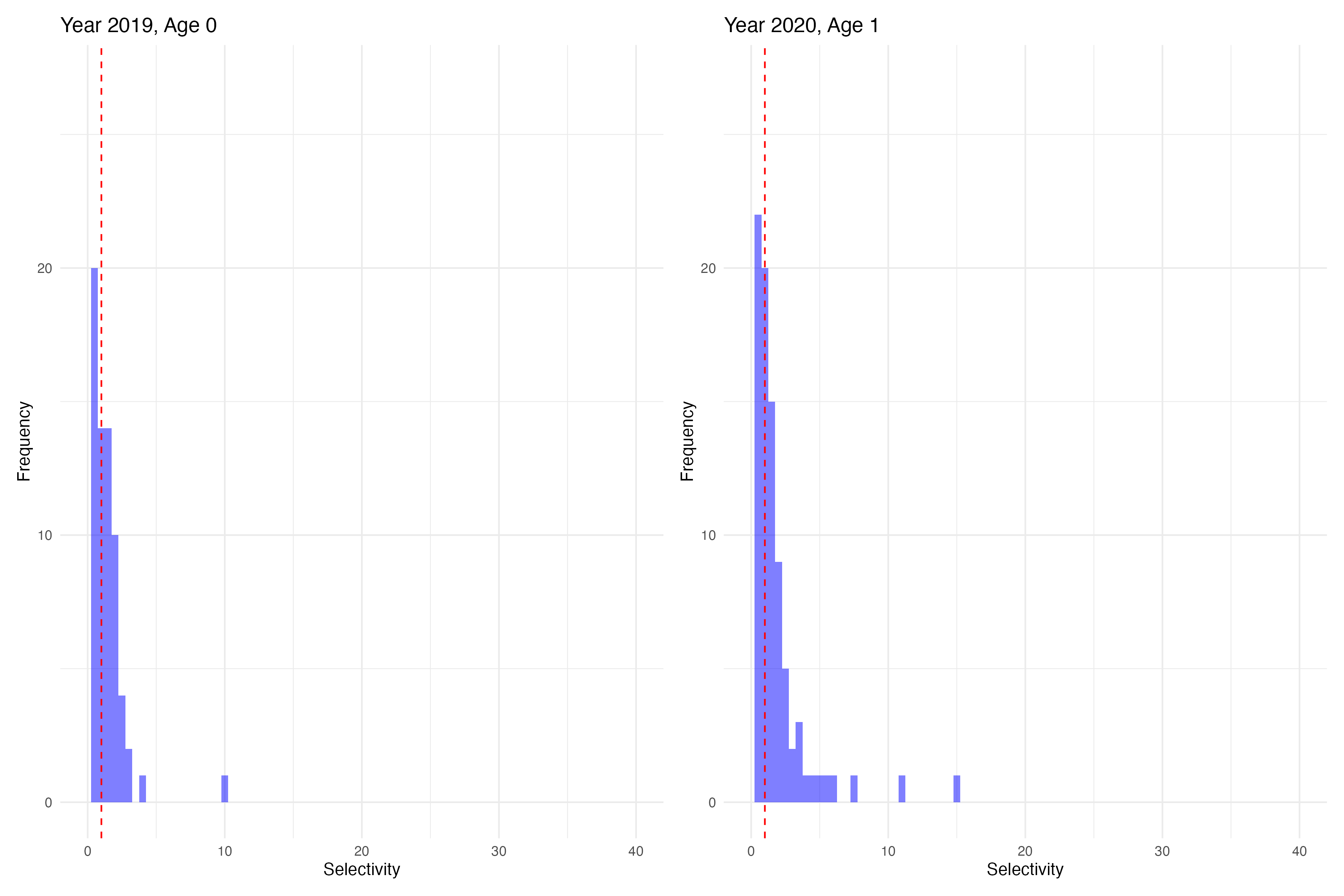}
    \caption{Histograms of Selectivity by Area (Year 2020 Age 0)}
    \label{fig:selectivity_histogram}
    \medskip 
    \begin{minipage}{0.65\textwidth} 
        {\scriptsize \textit{Note:} These are histograms of selectivity for daycare centers, focusing on age 0 applications in 2019 (left) and age 1 applications in 2020 (right). Selectivity is defined as the ratio of top-ranked applications to the number of vacant seats. The red vertical dotted line marks a selectivity of 1, representing an equal number of top-ranked applications and vacant seats.}
    \end{minipage}
\end{figure}

\noindent \textit{Signs of Vertical Differentiation.} 
Figure \ref{fig:selectivity_histogram} shows histograms of selectivity for daycare centers, focusing on age 0 applications in 2019 (left) and age 1 applications in 2020 (right). Selectivity is defined as the ratio of top-ranked applications to the number of vacant seats. The red vertical dotted line marks a selectivity of 1, representing an equal number of top-ranked applications and vacant seats. The histograms reveal substantial variation in selectivity across daycare centers. While many centers have a selectivity below 1, some exceed 10, indicating ten times more top-ranked applications than available seats for certain centers.

To examine the factors influencing daycare selectivity, I regress selectivity for each center-year-age combination on various daycare characteristics. These include the distance to the nearest station (in kilometers), the total number of full-time staff, total capacity, indicators for offering temporary childcare services and having undergone a third-party review, and an indicator for serving educational purposes as a certified facility combining daycare and education. The regression also includes fixed effects for the nearest station, year, and age. The OLS estimates, excluding the nearest station fixed effects, are presented in Table \ref{tab:selectivityregression}. The results are intuitive: daycare centers closer to the nearest station, employing more full-time staff, offering temporary childcare services, and providing educational services are more selective. Although the coefficient for third-party reviews is negative, it is not statistically significant.

\begin{table}[htbp!] \centering 
  \caption{Regression Results} 
  \label{tab:selectivityregression} 

    \begin{threeparttable}

\begin{tabular}{@{\extracolsep{5pt}}lc} 
\\[-1.8ex]\hline 
 & \multicolumn{1}{c}{\textit{Dependent variable:}} \\ 
\cline{2-2} 
\\[-1.8ex] & Selectivity \\ 
\hline \\[-1.8ex] 
 Distance from the Closest Station & $-$0.049 \\ 
  & (0.020) \\ 
 Total Number of Full-Time Staff & 0.063 \\ 
  & (0.025) \\ 
 Total Capacity & 0.002 \\ 
  & (0.004) \\ 
 Temporary Childcare Service & 0.770 \\ 
  & (0.198) \\ 
 Third Party Review & $-$0.051 \\ 
  & (0.119) \\ 
 Education and Care Preschool & 18.210 \\ 
  & (0.640) \\ 
  \hline
  Age Fixed Effects & Yes \\ 
Year Fixed Effects & Yes \\ 
Closest Station Fixed Effects & Yes \\ 
\hline \\[-1.8ex] 
Observations & 1,106 \\ 
R$^{2}$ & 0.561 \\ 
Adjusted R$^{2}$ & 0.548 \\ 
Residual Std. Error & 1.634 (df = 1073) \\ 
F Statistic & 42.817 (df = 32; 1073) \\ 
\hline \\[-1.8ex] 
\end{tabular} 

\begin{tablenotes}
    \scriptsize
    \item \textit{Note:} The unit of observation is center-year-age.
"Selectivity" is defined as the ratio of top-ranked applications to the number of vacant seats.
"Distance from the Closest Station" is measured in kilometers.
"Total Number of Full-Time Staff" denotes the total count of full-time employees at the center.
"Total Capacity" represents the maximum number of children the center can accommodate.
"Temporary Childcare Service" is a binary variable equal to 1 if the center offers short-term childcare services.
"Third Party Review" is a binary variable equal to 1 if the center has undergone an external evaluation.
"Education and Care Preschool" is a binary variable equal to 1 if the center is certified as a facility that integrates daycare and educational functions.
The regression includes fixed effects for the child’s age, year of observation, and the closest station to the center.
Standard errors are reported in parentheses.
\end{tablenotes}
        
    \end{threeparttable}
  
\end{table} 

\begin{figure}
    \centering
    \includegraphics[width=0.5\linewidth]{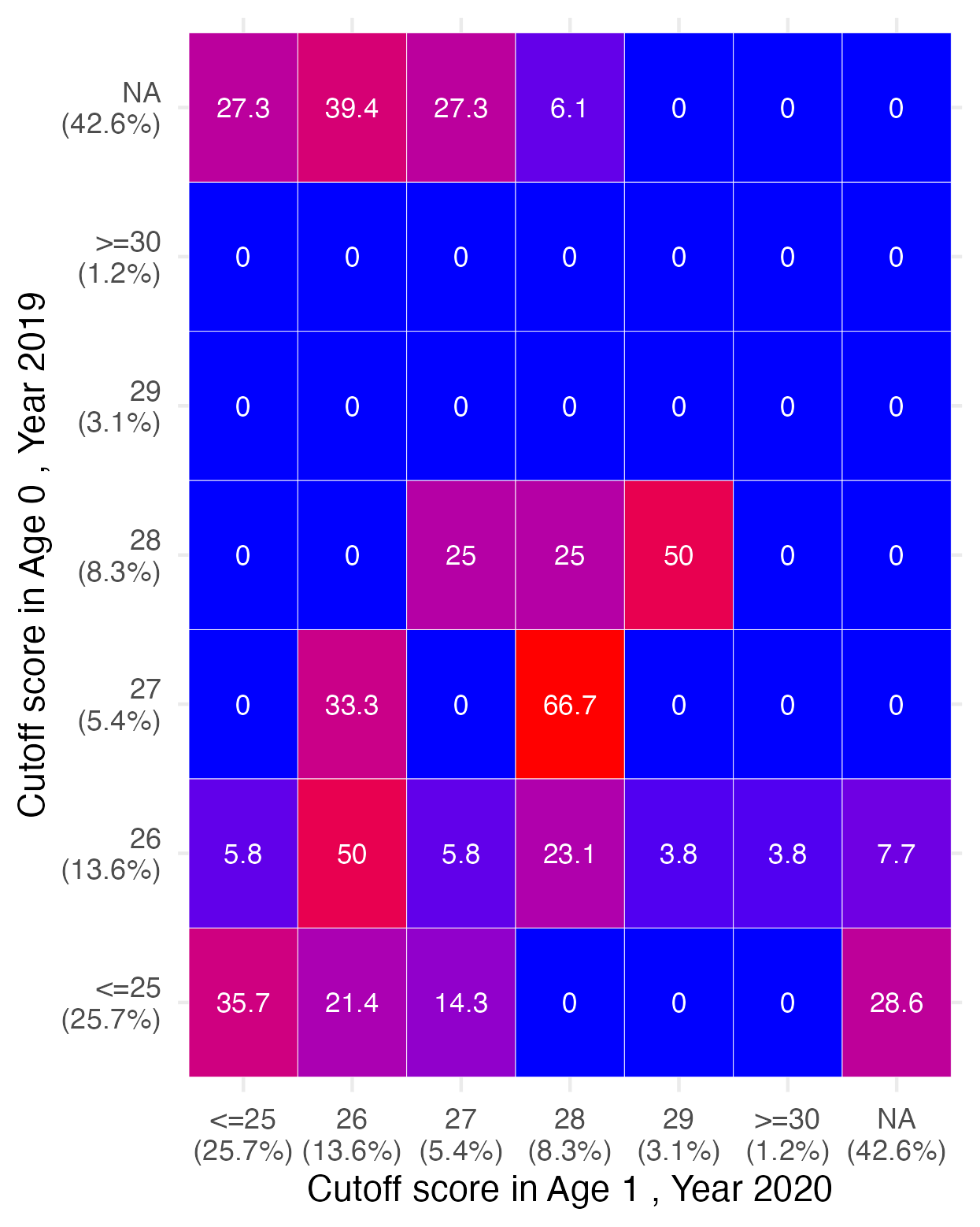}
    \caption{Transition Matrix of Cutoff Scores}
    \label{fig:score_transition}
        \begin{minipage}{0.8\textwidth} 
        {\scriptsize \textit{Note:} 
        The rows of the transition matrix represent the cutoff scores for year 2019, age 0, while the columns represent the cutoff scores for year 2020, age 1. Each cell shows the fraction of daycare centers with a specific cutoff value in age 1, conditional on having a specific cutoff value in age 0. Values in parentheses indicate the overall fraction of daycare centers with a given cutoff value in age 0 or age 1.
        \par}
    \end{minipage}
\end{figure}

\noindent \textit{Two points can make a big difference.} 
To illustrate the value of a two-point priority advantage on the waitlist, Figure \ref{fig:score_transition} presents a transition matrix of cutoff scores. Rows represent the cutoff scores for year 2019, age 0, while columns represent the cutoff scores for year 2020, age 1. Each cell shows the fraction of daycare centers with a particular cutoff value in age 1, conditional on having a specific cutoff value in age 0. The values in parentheses indicate the overall fraction of daycare centers with a given cutoff value in age 0 or age 1. For instance, 8.3\% of daycare centers had a cutoff of 28 in age 0, of which 50\% had a cutoff of 29 in age 1.

The matrix highlights the substantial potential benefit of being waitlisted and gaining two additional points. For example, consider an applicant with an initial score of 26 who was rejected by a daycare with a cutoff of 27. If the applicant chooses to remain on the waitlist instead of enrolling in a safety option, her score will increase to 28 the following year. Conversely, if she enrolls in a safety option and later reapplies, she will only have a score of 25 due to the penalty associated with transferring. Crucially, all daycares with a cutoff of 27 in age 0 have a cutoff of at most 28 and at least 26 in age 1. Thus, if the applicant strongly prefers a daycare she originally applied to, waiting an extra year could significantly improve her chances of acceptance. A similar argument applies to an applicant with an initial score of 27 applying to a daycare with a cutoff of 28 in age 0.

\noindent \textit{Being waitlisted and reapplying is common. Transfers are less frequent among younger applicants.} Tracking a fixed cohort across years highlights how common it is to be waitlisted and reapply. Among 732 age 0 applicants in 2019, 195 (26.6\%) were waitlisted, and of those, 131 (67.2\%) reapplied the following year. A similar pattern is seen from age 0 to age 1. Applicants typically apply at most twice, usually in consecutive years. For instance, fewer than 3\% of age 0 applicants in 2019 applied three times, and 90\% of those who applied twice did so in consecutive years. On the other hand, some applicants do reapply to transfer, but this is uncommon, especially for younger ages. For instance, only 4.7\% of age 1 applicants in 2020 were transfers. My structural model will account for reapplication only if the applicant was waitlisted in their first attempt the previous year, and will not allow for transfers.

\section{Preliminary Evidence}
\subsection{Documentation of Strategic Waiting}  \label{sec:preliminary_evidence}

In Section \ref{sec:illustrative_example}, I argued that prioritizing waitlisted applicants can incentivize them to initially refrain from listing their safety options, aiming to increase their chances of being admitted to more selective centers when they reapply. In this section, I will demonstrate that a substantial fraction of applicants submitted ROLs consistent with this strategic behavior in the actual data. To formalize this pattern, let $i = 1, 2, \dotsb, I$ index applicants, $j = 1, 2, \dotsb, J$ index centers, $t = 2019, 2020, 2021$ index years, and $a = 0, 1, \dotsb, 5$ index the age of the child. Suppose applicant $i$ initially applies in year $t(i)$ at age $a(i)$. Let $e_j^{1}$, $s_i^{1}$, and $R_i^{1}$ denote the cutoff score of center $j$, the priority score of applicant $i$, and the ROL of applicant $i$, respectively, in year $t(i)$ and age $a(i)$. Similarly, define $e_j^{2}$, $s_i^{2}$, and $R_i^{2}$ as the cutoff score of center $j$, the priority score of applicant $i$, and the ROL of applicant $i$, respectively, in year $t(i)+1$ and age $a(i)+1$. Then:

\begin{definition}[Dropping a Safety Option] \label{def:dropsafety}
The indicator variable ${DropSafety}_{i}$ equals one if and only if applicant $i$ \textit{drops a safety option}, meaning there exists a center $j$ such that: (i) $e_j^{1} \leq s_i^{1}$, (ii) $j \notin R_i^{1}$, and (iii) $j \in R_i^{2}$.
\end{definition}

\noindent In other words, an applicant drops a safety option if she lists a daycare center for the first time when reapplying, and that center had a cutoff score weakly smaller than her initial priority score.

The overall fraction of already waitlisted applicants who dropped a safety option after reapplying is 30\%, which is notably high. Table \ref{tab:postpone} breaks down this fraction by the initial age at the time of application. The table shows that the likelihood of reapplying and subsequently dropping a safety option generally decreases as the initial age of the applicant increases. The only exceptions are slight deviations observed between initial ages 1 and 2, and between ages 3 and 4, in the fraction of those dropping a safety option. This trend is intuitive, as older applicants, who have less time remaining until graduation, have less to gain from waiting an additional year to attend a more selective center compared to younger applicants.

\begin{table}[htbp!]

\begin{threeparttable}
\begin{tabular}{ccccc}
  \hline
Initial age & Number of
applications & Waitlisted & Reapplied & \textit{DropSafety} \\ 
  \hline
0 & 2026 & 350 (17.28\%) & 185 (52.86\%) & 66 (35.68\%) \\ 
  1 & 1929 & 485 (25.14\%) & 171 (35.26\%) & 48 (28.07\%) \\ 
  2 & 705 & 152 (21.56\%) & 48 (31.58\%) & 14 (29.17\%) \\ 
  3 & 579 & 66 (11.4\%) & 16 (24.24\%) & 0 (0\%) \\ 
  4 & 311 & 42 (13.5\%) & 10 (23.81\%) & 1 (10\%) \\ 
  All ages & 5550 & 1095 (19.73\%) & 430 (39.27\%) & 129 (30\%) \\ 
   \hline
\end{tabular}

\begin{tablenotes}
\scriptsize
\item \textit{Note:} The table summarizes the fraction of applicants who dropped a safety option (see \ref{def:dropsafety} for definition) after reapplying, broken down by their initial age at the time of application. The columns display the total number of applications, the number and percentage of applicants who were waitlisted, the number and percentage of waitlisted applicants who reapplied, and the number and percentage of those who dropped a safety option. Percentages in parentheses are calculated relative to the values in the previous column. The row "All ages" aggregates the data across all initial ages.
\end{tablenotes}

\end{threeparttable}

\caption{Fraction of Applicants Dropping a Safety Option}
\label{tab:postpone}

\end{table}

\subsection{Benefit from Being Waitlisted}
In the previous subsection, I demonstrated the prevalence of applicants dropping safety options, a behavior that can result from strategic efforts to become waitlisted. To confirm that this pattern is driven by dynamic incentives rather than other factors, such as a declining value of outside options, I will show in this subsection that applicants who stand to benefit more from being waitlisted are indeed more likely to drop their safety options.

\begin{figure}[htbp]
    \centering
    \begin{subfigure}[b]{0.8\textwidth}
        \centering
        \includegraphics[width=\textwidth]{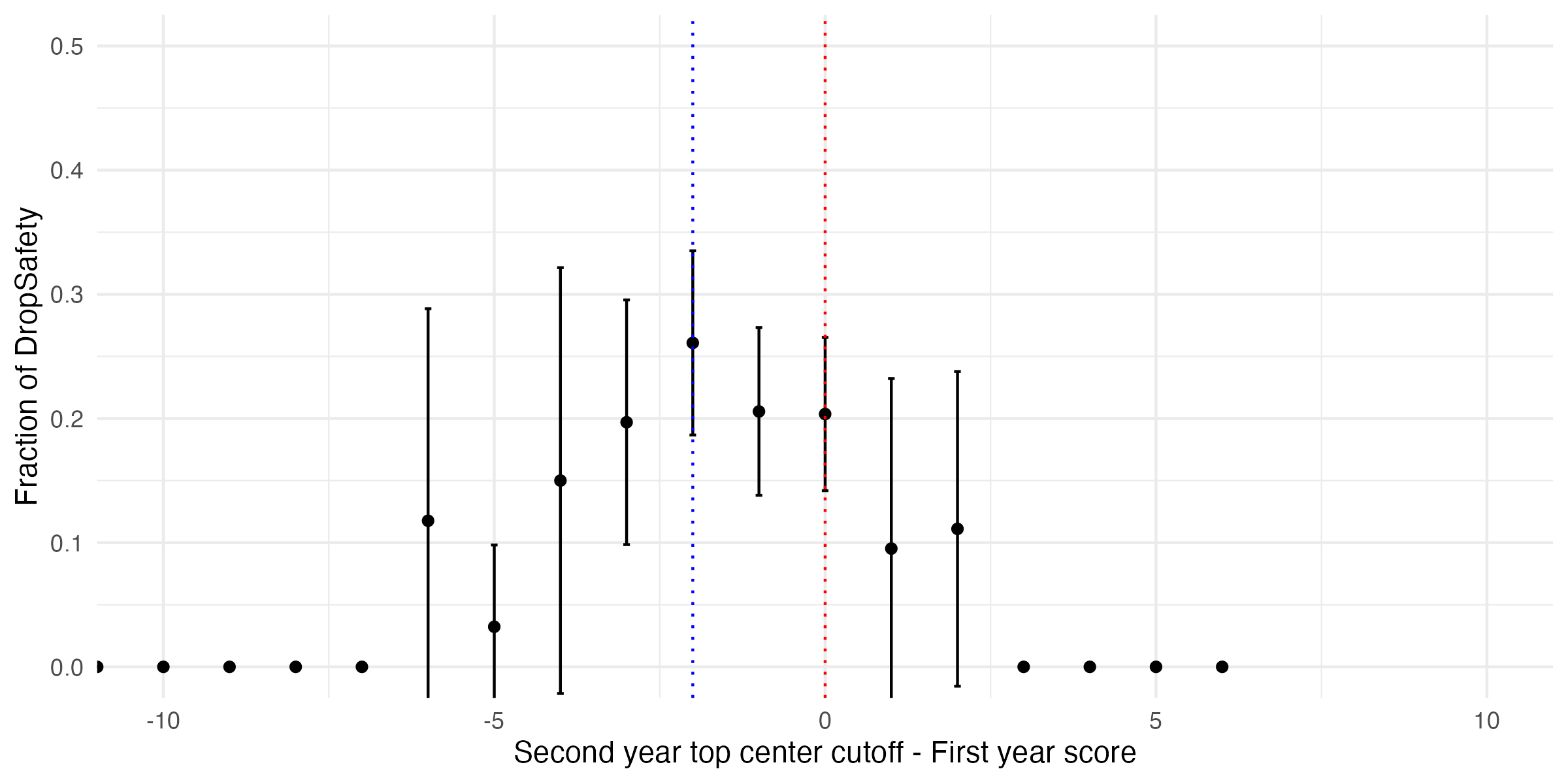}
        \caption{Fraction of \textit{DropSafety} by Distance to Cutoff of Second Round Top Choice}
        \label{fig:psc_a}
    \end{subfigure}
    \vspace{0.5cm} 
    \begin{subfigure}[b]{0.8\textwidth}
        \centering
        \includegraphics[width=\textwidth]{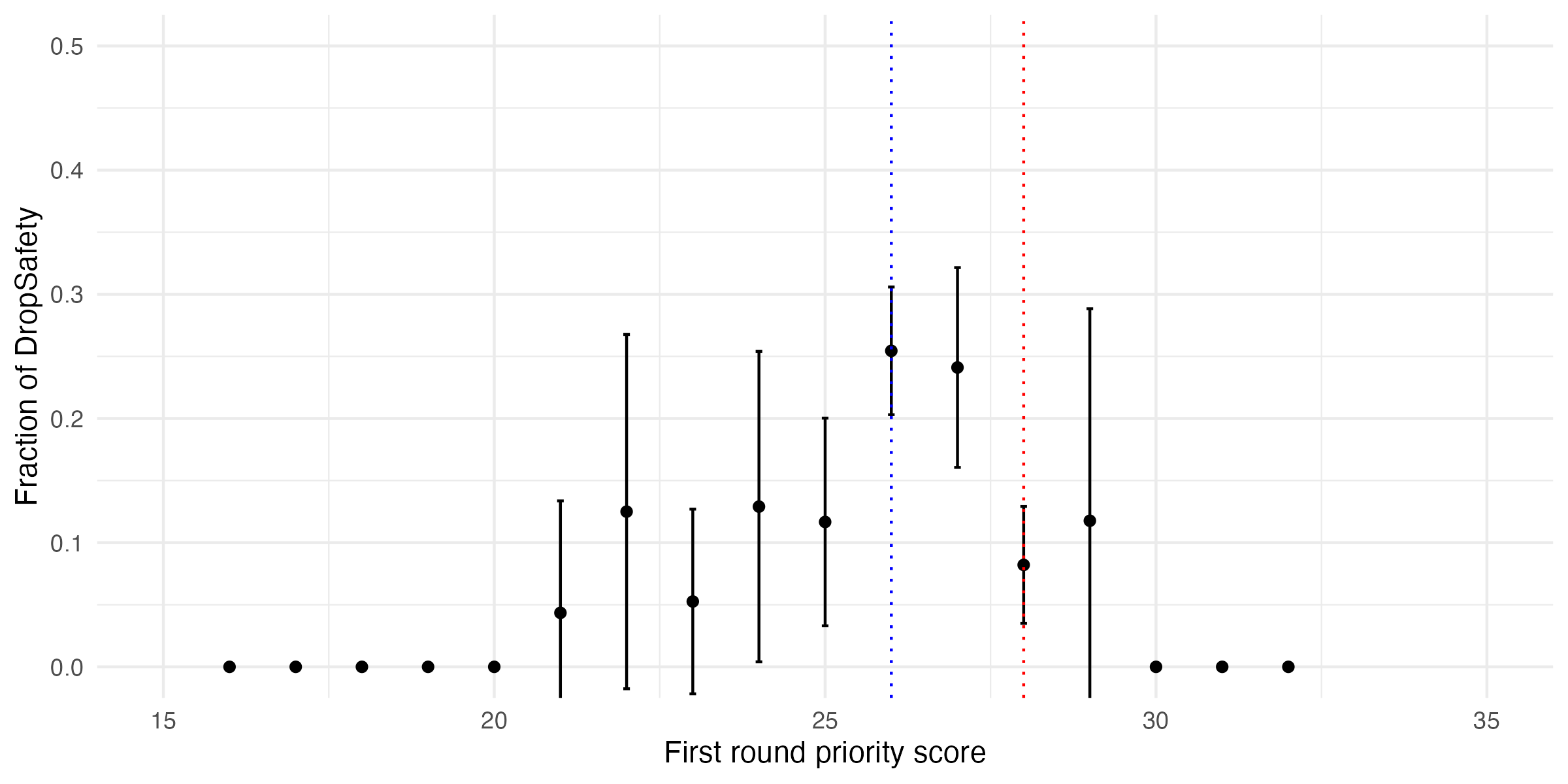} 
        \caption{Fraction of \textit{DropSafety} by Initial Priority Score}
        \label{fig:psc_b}
    \end{subfigure}
    \caption{Fraction of \textit{DropSafety} by Priority Status}
    \medskip 
    \begin{minipage}{0.8\textwidth} 
        {\scriptsize \textit{Note:} Panel A shows the Fraction of already waitlisted reapplicants who dropped their safety options based on the number of additional points needed to reach the cutoff of their top-choice daycare center in the second round of applications. The red dotted line at 0 indicates no additional points needed, while the blue dotted line at -2 indicates that two additional points were needed. Panel B presents a similar analysis using the applicant's initial priority score as the running variable. The red dotted line at 28 marks the typical cutoff score of a selective and popular daycare, and the blue dotted line at 26 represents the typical score of an average applicant. Vertical lines correspond to 95\% confidence intervals.
        \par}
    \end{minipage}
    \label{fig:psc}
\end{figure}

Figure \ref{fig:psc} presents two panels that analyze the behavior of waitlisted reapplicants who dropped their safety options. Panel A breaks down the share of these applicants based on the number of additional points needed to reach the cutoff of their top choice. The horizontal axis represents the difference between the cutoff score of the first-ranked daycare center in the second round and the applicant's initial priority score, with the red dotted line at 0 indicating no additional points needed and the blue dotted line at -2 indicating that two additional points were needed. The vertical axis shows the fraction of applicants who dropped their safety options for each value of this difference. Panel B offers a similar analysis but uses the applicant’s initial priority score as the running variable, with the red dotted line at 28 marking the typical cutoff score of a selective and popular daycare, and the blue dotted line at 26 representing the typical score of an average applicant. The figure supports the expectation that applicants needing exactly two additional points to reach the cutoff of their top choice in the second round (as shown in Panel A, 26.1\%) or those with an initial score close to the typical selective cutoff of 28 (as shown in Panel B, 25.4\%) are most likely to drop their safety options.

To further investigate whether the relationship between being waitlisted and dropping safety options persists after accounting for age and other factors, I perform a regression analysis. As discussed in Section \ref{sec:illustrative_example}, the benefit of being waitlisted is realized only if it increases the likelihood of an applicant being admitted to their preferred daycare. To measure this benefit, I define a variable $\Delta_i^k$ that captures whether being waitlisted improves an applicant’s chances of meeting a certain admission threshold.

\begin{equation}
    \Delta_i^k = \mathds{1}\{s_i^{1} < \bar{s}^{1,k} \text{~and~} s_i^{1} + 2 \geq \bar{s}^{2,k} \} \label{eq:benefit}
\end{equation}

This variable, \(\Delta_i^k\), equals 1 if being waitlisted allows applicant \(i\) to improve their priority score to meet or exceed a specified threshold. The thresholds \(\bar{s}^{1,k}\) and \(\bar{s}^{2,k}\) can be defined in several ways, with \(k\) indexing one of the following specifications: (i) Actual Cutoffs: \(\bar{s}^{1,k}\) is the actual cutoff of the highest-ranked daycare in \(R^1_i\), and \(\bar{s}^{2,k}\) is the cutoff of the highest-ranked daycare in \(R^2_i\); (ii) Fixed Cutoff of 28: Both \(\bar{s}^{1,k}\) and \(\bar{s}^{2,k}\) are set to 28, representing the typical cutoff score of a highly selective daycare\footnote{I choose this value since it is argued to be the typical cutoff score of popular, selective daycare centers.
In fact, \textit{Hokatsu Manual}, a website that gives advice on how to plan application strategies, emphasizes that admission to popular centers are "competition between score 28 applicants".
{See "2019 Bunkyo Ward Daycare Admission Results," \textit{Hokatsu Megurokko}, accessed July 4, 2024, \url{https://hokatsu.megurokko.com/bunkyouku-kekka2019/}.}}; (iii) 90th Percentile: \(\bar{s}^{1,k}\) and \(\bar{s}^{2,k}\) are set to the 90th percentile of cutoffs for year-age pairs \((t(i), a(i))\) and \((t(i)+1, a(i)+1)\), reflecting variations in selectivity; and (iv) Modal Cutoffs: \(\bar{s}^{1,k}\) and \(\bar{s}^{2,k}\) are defined as the most common cutoff scores (mode) for year-age pairs \((t(i), a(i))\) and \((t(i)+1, a(i)+1)\).


Using these definitions, I estimate a linear regression model where \({DropSafety}_i\) is regressed on \(\Delta_i^k\), with fixed effects for the applicant’s initial age, region, and year of application. I use the subsample of waitlisted reapplicants for this analysis. Table \ref{tab:regress} presents the OLS estimates, with each $k$th column corresponding to a specification using $\Delta_i^k$ and the coefficient of interest being $\beta^k$. In column (1), the coefficient $\beta^1$ is estimated to be positive and statistically significant, as expected. Specifically, an applicant whose priority score can reach the cutoff of their second-round top choice with two extra points is 8.1 percentage points more likely to drop a safety option than those who cannot. The coefficients $\beta^2$ and $\beta^3$ are also positive and statistically significant, with somewhat larger magnitudes than $\beta^1$. In contrast, $\beta^4$ is imprecisely estimated with a negative point estimate. Since $\Delta^4$ is based on the modal cutoff, this suggests that manipulation of priority scores by altering ROLs is more common among middle-score applicants who aim for more selective daycare centers, rather than among low-score applicants who are more likely to be waitlisted regardless of their choice of ROLs.
The regression results also provide evidence against the concern that the observed pattern of dropping safety options is solely driven by the declining value of the outside option with age, since all specifications include age fixed effects. 

\begin{table}[htbp!] \centering 
  \caption{Estimation results} 
  \label{tab:regress} 

\begin{threeparttable}
    \begin{tabular}{@{\extracolsep{5pt}}lcccc} 
\\[-1.8ex]\hline 
 & \multicolumn{4}{c}{\textit{Dependent variable:}} \\ 
\cline{2-5} 
\\[-1.8ex] & \multicolumn{4}{c}{DropSafety} \\ 
\\[-1.8ex] & (1) & (2) & (3) & (4)\\ 
\hline \\[-1.8ex] 
 $\Delta^1$ & 0.083 &  &  &  \\ 
  & (0.030) &  &  &  \\ 
 $\Delta^2$ &  & 0.137 &  &  \\ 
  &  & (0.029) &  &  \\ 
 $\Delta^3$ &  &  & 0.107 &  \\ 
  &  &  & (0.029) &  \\ 
 $\Delta^4$ &  &  &  & $-$0.049 \\ 
  &  &  &  & (0.040) \\ 
\hline \\[-1.8ex] 
Age Fixed Effects & Yes & Yes & Yes & Yes \\ 
Year Fixed Effects & Yes & Yes & Yes & Yes \\ 
Area Fixed Effects & Yes & Yes & Yes & Yes \\ 
\hline 
Observations & 671 & 671 & 671 & 671 \\ 
R$^{2}$ & 0.073 & 0.093 & 0.081 & 0.064 \\ 
Adjusted R$^{2}$ & 0.055 & 0.075 & 0.063 & 0.045 \\ 
Residual Std. Error (df = 657) & 0.370 & 0.367 & 0.369 & 0.372 \\ 
F Statistic (df = 13; 657) & 3.981 & 5.153 & 4.450 & 3.450 \\ 
\hline 
\end{tabular} 

\begin{tablenotes}
    \scriptsize
    \item \textit{Note:} The table presents the OLS regression results for the relationship between \({DropSafety}_i\) (see \ref{def:dropsafety} for definition) and \(\Delta_i^k\) (see Equation \ref{eq:benefit} for definition), under four different specifications (\(\Delta^1, \Delta^2, \Delta^3, \Delta^4\)). Each column corresponds to a different specification for \(\Delta^k\). The table includes fixed effects for the applicant’s initial age, region, and year of application. Standard errors are reported in parentheses. 
\end{tablenotes}
\end{threeparttable}

\end{table} 

\section{Model}
To build and estimate a structural model of daycare center choice that incorporates waitlisting and reapplication, I build upon \cite{agarwalDemandAnalysisUsing2018}'s static framework of school choice, which incorporates strategic behavior of applicants. 
I extend their framework by allowing applicants to become waitlisted, after which they can reapply in the subsequent year. 
An applicant is indexed by $i \in \mathcal{I}$, a daycare center by $j \in \mathcal{J}$. 
An applicant enters the market at an exogenously given age $a_0$, and exits after age 5.
Each applicant faces two periods $t = 1,2$, where period 1 corresponds to the year when the applicant's age is $a_0$, and period 2 corresponds to the years during ages $a_0 + 1, a_0 + 2, \dotsb, 5$.

\subsection{Preferences} \label{sec:preferences}
Applicant $i$ attending daycare center $j$ receives the following privately observed, constant flow utility at each age:
\[
    v_{ij} = \alpha_j + \beta_j s_{i}^{1} + \gamma d_{ij} + \epsilon_{ij}
\]
Here, $\alpha_j$ represents daycare fixed effects that capture vertical differentiation. $s_{i}^{1}$ denotes applicant $i$'s priority score in her initial round of application, which captures her necessity of using an accredited daycare center. I allow its effect to vary according to the daycare center, as captured by $\beta_j$. The variable $d_{ij}$ is an indicator that equals 1 if applicant $i$ and center $j$ are in the same area, and 0 otherwise. $\gamma$ measures the disutility of commuting to a daycare center outside the applicant's area of residence. For scale normalization, I set $\gamma = -1$. $\epsilon_{ij}$ is an unobserved utility shock that is normally distributed and independent across applicants: $\epsilon_i = (\epsilon_{i1}, \dotsb, \epsilon_{iJ}) \sim \mathcal{N}(0, \Sigma)$. I assume that $\epsilon_i \perp (s_i^1, d_{ij})$. This implies that the unobserved, heterogeneous part of applicant preference is independent from the parent's residential choice and labor supply status, as the latter is included in $s_i^1$. While I assume that the flow utility $v_{ij}$ is constant across ages, I allow the value of the outside option to vary with age and normally distributed:
\[
    v_{i0}^{a} \sim \mathcal{N}(\mu_0^a, {\sigma_0^{a}}^2)
\]
I assume that the applicant has perfect foresight on the exact values of $\{v_{i0}^{a}\}_{a=a_0}^{5}$ at the beginning of period 1. For location normalization, I set $\mu_0^1 = 0$. I denote applicant $i$’s yearly discount factor as ${\delta}$ and set this value to 0.95.

\subsection{Assignment Mechanism} \label{sec:assmech}  
The assignment mechanism determines the allocation of applicants based on their submitted ROLs and priority scores for a given year and age. Suppose each applicant $i = 1, 2, \dotsb, N$ submits a ROL $R_i \in \mathcal{R}$ and has a priority score $s_i \in \mathcal{S} = [\underline{s}, \underline{s} + 1, \dotsb, \overline{s}]$, where $\mathcal{R}$ is a set of ordered lists with up to 5 choices from $\mathcal{J}$. The centralized matching function $\Phi: \mathcal{R}^N \times \mathcal{S}^N \rightarrow (\{0,1\})^{JN}$ determines each applicant’s allocation using a truncated serial dictatorship algorithm. For each applicant $i$, $\Phi$ returns a $J$-dimensional vector $(\{0,1\})^J$ where the $j$-th element is 1 if and only if she is assigned to center $j$. If all elements of this vector are zero, it means the applicant is waitlisted. 
The mechanism calculates a unique cutoff score $e_j$ for each center $j$, below which applicants will not be admitted. Applicant $i$ is assigned to the highest-ranked center on her ROL where the cutoff score does not exceed her priority score.

Applicant \( i \) selects \( R^1_i \) from \( \mathcal{R} \). If she is waitlisted in period 1, her priority score increases by 2 points (\( s_i^2 = s_i^1 + 2 \)), and she has the option to submit another ROL, \( R_i^2 \in \mathcal{R} \cup \{\emptyset\} \), in period 2. Here, with a slight abuse of notation, I have introduced the possibility for the reapplicant to choose not to reapply by allowing her to submit an empty ROL, \( \emptyset \). If she is assigned in period 1, she is not permitted to reapply, and her assignment remains the same in period 2.

\subsection{Applicants' Problem}

Given a belief over the distribution of cutoffs and her own priority score $s_i^t$, by submitting a ROL $R_i^t$ applicant $i$ faces a lottery $L^t(R^t_i, s^t_i) \in \Delta^{\mathcal{J}}$ whose $j$th component corresponds to the believed assignment probability to center $j$. In my baseline model, I assume that each applicant has correct beliefs over the current and future distribution of cutoffs given her priority score:
\begin{assumption} \label{ass:ratex}
    At the beginning of period 1, applicant $i$ has rational expectations over the distribution of cutoffs both in periods 1 and 2 given her own score $s_i^1$.
\end{assumption}
\noindent Given this, the lottery applicant $i$ faces in period $t=1,2$ can be expressed as:
\begin{equation}
    L^t(R_i^t, s_i^t) = \mathds{E}[\Phi((R^t_i, s^t_i), (R^t_{-i}, s^t_{-i})) | R_i^t, s_i^t]
\end{equation}
where the expectation is taken over the true distribution of other applicants' priority scores and preferences, as well as their strategies on submitting ROLs. Note that Assumption \ref{ass:ratex} implies that applicant $i$ also knows in period 1 the object $L^2(\cdot, s_i^{2})$, which returns the lottery she faces in period 2 for a given ROL. 
Note that for a given lottery $L^t(R^t_i, s^t_i)$, the probability of being waitlisted, which I denote as $p^t(R^1_i, s^1_i)$, can be expressed as: $p^t(R^1_i, s^1_i) = 1 - \iota \cdot L^t(R^t_i, s^t_i)$ where $\iota$ is a $|\mathcal{J}|$ dimensional vector of ones. 
In choosing the pair $(R_i^1, R_i^2)$, the applicant solves the following problem:

\begin{align*}
    \max_{R_i^1 \in \mathcal{R}, R_i^2 \in \mathcal{R} \cup \{0\}} &\Bigg\{ (1 - p^1)
    \underbrace{\Bigg[\frac{v_i \cdot L^1}{1 - p^1} + \delta \frac{v_i \cdot L^1}{1 - p^1} + \dotsb + \delta^{5 - a_0} \frac{v_i \cdot L^1}{1 - p^1}\Bigg]}_{\text{Not waitlisted in period 1}} \\
    &+ p^1 \Bigg[ (1 - p^2) \underbrace{\Big( v_{i0}^{a_0} + \delta \frac{v_i \cdot L^2}{1 - p^2} + \delta^2 \frac{v_i \cdot L^2}{1 - p^2} + \dotsb + \delta^{5 - a_0} \frac{v_i \cdot L^2}{1 - p^2} \Big)}_{\text{Waitlisted in period 1, not waitlisted in period 2}} \\
    &\quad + p^2 \underbrace{\Big( v_{i0}^{a_0} + \delta v_{i0}^{a_0 + 1} + \dotsb + \delta^{5 - a_0} v_{i0}^{5} \Big)}_{\text{Waitlisted in both period 1 and period 2}} \Bigg] \Bigg\}
\end{align*}
\text{where}
\begin{align*}
    L^1 &= L^1(R^1_i, s^1_i) \\
    p^1 &= 1 - \iota \cdot L^1 \\
    L^2 &= L^2(R^2_i, s^1_i + 2) \\
    p^2 &= 1 - \iota \cdot L^2
\end{align*}
The first term represents the expected total utility when the applicant is not waitlisted in period 1. In this case, the applicant remains at the assigned daycare until graduation. The second term corresponds to the scenario where the applicant is waitlisted in period 1 but not in period 2. Here, she receives utility from the outside option, \( v_{i0}^{a_0} \), during period 1, and then gets assigned to a daycare in period 2, staying there until graduation. The last term accounts for the situation where the applicant is waitlisted in both periods 1 and 2, in which case she continues to use the outside option each period until graduation.

The applicant's problem can be succinctly rewritten as follows:
\begin{equation} \label{eq:obj}
    \max_{R_i^1, R_i^2 \in \mathcal{R}} \Bigg\{
    v_i \cdot \Tilde{L} 
    + \left( v^{a_0}_{i0} p^1 + \delta v^{a_0+1}_{i0} p^2 + \dotsb + \delta^{5 - a_0} v^5_{i0}p^2 \right)
    \Bigg\}
\end{equation}

\noindent where%
\begin{align}
    \Tilde{L} &= (1 - \Tilde{p}) L^1 + \Tilde{p} L^2 \label{eq:ltilde} \\
    \Tilde{p} &= \frac{\Tilde{\delta}}{1 + \Tilde{\delta}} p^1 \\
    \Tilde{\delta} &= \frac{{\delta}(1 - {\delta}^{5-{a_0}})}{1-{\delta}}
\end{align}
In essence, the applicant's problem boils down to choosing a single lottery \( \Tilde{L} = \Tilde{L}(R^1_i, R^2_i, s^1_i) \), determined by the two ROLs \( R^1_i \) and \( R^2_i \), while considering the value of the outside option. The first term of the objective represents the expected total utility from being assigned to an accredited daycare, while the second term captures the expected total utility from utilizing the outside option.


\section{Identification} \label{sec:identification}
My goal is to identify the distribution of $v_i$'s parametrized by $\theta = (\alpha, \beta, \Sigma)$. For now I assume that the value of the outside option is always zero. First, note that applicant $i$'s observed choice of $R^1_i$ and $R^2_i$ defines a region $\Tilde{C}(R^1_i, R^2_i)$ such that $v_i$ should lie in:
\begin{equation}
    \Tilde{C}(R^1, R^2) = \left\{v \in \mathbb{R}^{J}: v \cdot\left(\Tilde{L}(R^1, R^2)-\Tilde{L}'\right) \geq 0, ~\forall \Tilde{L}' \in \Tilde{\mathcal{L}}\right\}
\end{equation}
where I dropped the subscript $i$ for notational convenience. Hence the probability of choosing $(R^1, R^2)$ conditional on covariates $z_i = (s_i, \{d_{ij}\}_{j=1}^{J})$ can be expressed as follows:
\begin{align}
    \mathds{P}(R^1, R^2 \mid z) = \int \mathds{1}\left\{v \in \Tilde{C}(R^1, R^2)\right\} f_{V \mid z, \theta}(v \mid z, \theta) \mathrm{d} v \label{eq:likeliprob}
\end{align}
However, for an applicant who does not become waitlisted in period 1, I do not observe her $R^2$, the ROL she would have submitted in period 2 had she been waitlisted in period 1. For such applicant $i$, her $v_i$ lies in a region $C(R_i^1)$ defined as follows:
\begin{align}
    C(R^1) &= \bigcup_{R^2 \in \mathcal{R}}{\Tilde{C}(R^1, R^2)} \label{eq:unions} \\
    &= \left\{v \in \mathbb{R}^{J}: \exists R^2 \in \mathcal{R}~\forall \Tilde{L}' \in \Tilde{\mathcal{L}}~~~ v \cdot\left(\Tilde{L}(R^1, R^2)-\Tilde{L}'\right) \geq 0 \right\} \label{eq:cexist}
\end{align}
$C(R^1)$ is the set of utility vectors that rationalizes the observed choice of $R^1$.

\begin{figure}[htbp]
    \centering
    \includegraphics[width = 0.75 \textwidth]{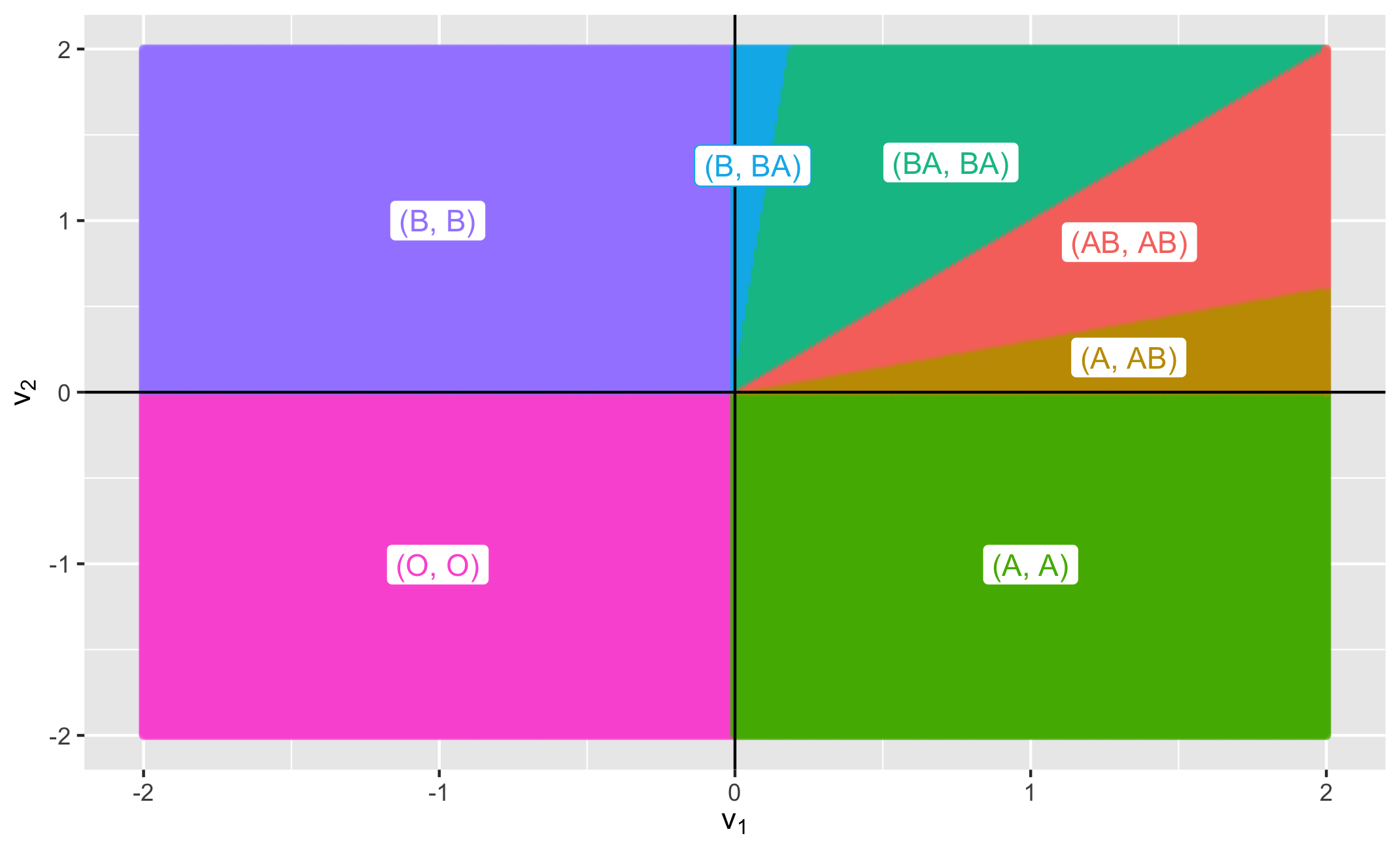}
    \caption{Partition of $\mathbb{R}^2$ by optimal $(R^1, R^2)$}
    \medskip 
    \begin{minipage}{0.65\textwidth} 
        {\scriptsize \textit{Note:} A simulation with $\pi_A^1 = 0.1$, $\pi_A^2 = 0.3$, $\pi_B^1 = 0.6$, $\pi_B^2 = 0.7$, $\Tilde{\delta}=2$ (for the definition of $\pi^t_j$'s, see Section \ref{sec:estimation}.). I discretize the interval $[-2,2]^2$ and obtain $(v_{iA}, v_{iB})$ for each applicant. Each applicant chooses $R^1$ and $R^2$, each from $\{AB, A, BA, B, O\}$ where $O$ corresponds to the outside option, to maximize her expected utility defined in equation (\ref{eq:obj}). Each coded area corresponds to the optimal pair of ROLs.
        \par}
    \end{minipage}
    \label{fig:simat}
\end{figure}

Figure (\ref{fig:simat}) shows how $\Tilde{C}(R^1, R^2)$ and $C(R^1)$ would look like, based on a simulation of a simplified model with $J=2$ schools $A$, $B$, and an outside option $O$. Each $\Tilde{C}(R^1, R^2)$ is corresponds to the region labelled $(R^1, R^2)$. For example, an applicant in region $(A, AB)$ prefers both schools $A$ and $B$, but it is optimal for her to not list $B$ in her first round, in order to become waitlisted and increase her chances of being admitted to school $A$ in her second round. $C(R^1)$ is constructed by taking unions of $\Tilde{C}(R^1, R^2)$ with the same $R^1$'s. For example, for an applicant who lists only school $A$ in her first period and does not get waitlisted, I know that her utility vector lies in the region defined as the union of region $(A, AB)$ and region $(A,A)$.

Since $\Tilde{C}(R^1, R^2)$ indexed by $(R^1, R^2)$ creates a partition of the space of utility vectors, $\mathds{R}^J$, so does $C(R^1)$. Hence the likelihood of observing $R^1$ can be expressed as follows:
\begin{align}
    \mathds{P}(R^1 \mid z) = \int \mathds{1}\left\{v \in C(R^1)\right\} f_{V \mid z, \theta}(v \mid z, \theta) \mathrm{d} v \label{eq:likelihood}
\end{align}
Note that the same pair of ROLs $(R^1, R^2)$ leads to different induced lotteries $\Tilde{L}(R^1, R^2)$, depending on the year $y \in \{2019, 2020, 2021\}$ and age $a \in \{0, 1, \dotsb, 5\}$ $R^1$ is submitted, even holding the priority score $s^1$ constant. If I assume that the distribution of $v$ does not depend on $y, a$ so that $v|z, y, a \sim v|z, y', a'$ for any $(y,a)$ and $(y',a')$, then variation in $(y,a)$ help me identify $\theta$. 
An identification for the outside option parameters $\mu^a$ and ${\sigma^a}^2$ follows a similar manner, that I observe the probability of an applicant not reapplying even after being waitlisted, conditional on her covariates. Then the yearly variation leads to an exogenous shift in this probability.


Due to limited support of $(y,a)$, I will only have partial identification.
In my application I assume that $v_{ij}$ and $v_i^a$ are normally distributed conditional on the covariates $z_{ij}$, and hence point identification of $\theta$ comes from this parametric assumption.

\section{Estimation} \label{sec:estimation}
Estimation takes two steps, where in the first stage I estimate the assignment probabilities and in the second stage I estimate preference parameters $\theta = (\beta, \Sigma)$.

\subsection{First Stage: Assignment Probabilities}  \label{sec:firststage}
\cite{agarwalDemandAnalysisUsing2018} showed that for mechanisms that belong to a class called "Report-Specific Priority + Cutoff (RSP+C)" mechanisms, assignment probabilities can be consistently estimated by resampling students. The truncated serial dictatorship algorithm used in Bunkyo municipality also belongs to this class, so I can use their bootstrap estimator, shown below:
\begin{align}
    \hat{L}^t(R_i^t, s_i^t) &= \frac{1}{B}\sum_{b=1}^{B} \Phi((R_i^t, s_i^t), (R_{-i}^t, s_{-i}^t)_b)
\end{align}
where $B$ is the number of bootstrap draws. 
For each bootstrap sample of applicants, I compute the cutoff for each school by simulating a truncated serial dictatorship algorithm. Given a distribution of cutoff $e^t_j$ for each school $j$, the probability of being assigned to school $j$ for an applicant with priority score $s^t$ who submits ROL only consisting of school $j$ is:
\begin{equation}
    \pi^t_j(s) = \mathbb{P}(e^t_j \leq s) \label{eq:pi}
\end{equation}
Since a student's priority is not specific to her ROL under the truncated serial dictatorship algorithm, assuming independence of the distribution of cutoffs of different centers, the probability of her being assigned to a particular school does not change based on where the school is listed in her ROL, conditional on being rejected by every school listed earlier. This feature can be used to simplify the estimation of $L^t(R,s)$. If I denote the $k$-th element of ROL $R$ as $R_k$, then the $R_k$th element of $L(R,s)$ is:
\begin{equation}
    L^t_{R_k}(R,s) =  \left( \prod_{k' < k}{(1-\pi^t_{R_{k'}})} \right) \pi^t_{R_k} \label{eq:lhat}
\end{equation}
$\hat{L}^t_{R_k}(R,s)$ can be obtained by plugging in estimates of $\pi^t_j$'s, which can be obtained from equation (\ref{eq:pi}) and the bootstrap distributions of $e_j$'s, to the right hand side of equation (\ref{eq:lhat}). Estimate of $\Tilde{L}(R^1_i, R^{2}_i, s^1_i)$ can then be obtained by substituting $\hat{L}^1(R_i^1, s_i^1)$, $\hat{L}^2(R_i^{2}, s_i^{2})$ and an estimate or a plugged in value for $\delta$ to the right hand side of equation (\ref{eq:obj}).

\subsection{Second Stage: Preference Parameters}
Given the first stage estimates of the lotteries, I estimate the preference parameter $\theta$ using method of simulated moments. Let $r = 1, \dotsb, R$ index targeted moments. I define the sample moment $\hat{h}_{a,y}^r(\theta)$, conditional on the applicant initially applying at age $a$ in year $y$, as follows:
\[
\hat{h}_{a,y}^r(\theta) = 
\frac{1}{|\mathcal{I}_{a,y}|}\sum_{i \in \mathcal{I}_{a,y}}{\left[
m_{i,r}
- \frac{1}{S}\sum_{s=1}^{S}{
\hat{m}_{i,r}^s(\theta)
}\right]}
\]
Here, $\mathcal{I}_{a,y}$ represents the set of applicants who initially applied at age $a$ in year $y$. The term $m_{i,r}$ refers to the $r$th individual moment, as detailed in Table \ref{tab:moments}. The number of simulation draws, denoted by $S$, is set to 100. The simulated individual moment is given by $\hat{m}_{i,r}^s(\theta)$. The simulation proceeds by first generating the flow utilities $v_{ij}$ and $v_{i0}^a$ for each applicant. Using these utilities, each applicant’s optimal ROLs are determined based on the first-stage lottery estimates. The serial dictatorship algorithm is then executed for each age group, sequentially across years, with center capacities updated accordingly at each step.

\begin{table}[htbp!]
\centering

\begin{threeparttable}
    \begin{tabular}{ll}
\toprule
{Individual Moment} & {Description} \\
\midrule
$\mathds{1}\{j \in R^1_i\}, j = 1, \dotsb, J$ & Lists a particular center in her initial ROL \\
$\mathds{1}\{j \in R^1_i\} s_i^1, j = 1, \dotsb, J$ & Above interacted with initial priority score \\
$\mathds{1}\{\iota \cdot \lambda^1_i = 0\}$ & Waitlisted in period 1 \\
$\mathds{1}\{j \in R^2_i\}, j = 1, \dotsb, J$ & Lists a particular center in her second ROL \\
$\mathds{1}\{j \in R^2_i\} s_i^1, j = 1, \dotsb, J$ & Above interacted with initial priority score \\
$\mathds{1}\{\iota \cdot \lambda^2_i = 0\}$ & Waitlisted in period 2 \\
$DropSafety_{i}$ & DropSafety in period 1 \\
$\mathds{1}\{R^2_i = \emptyset\}$ & Does not reapply in period 2 \\
\bottomrule
\end{tabular}

\begin{tablenotes}
    \scriptsize \item \textit{Note:} The table lists the individual moments \(m_{i,r}\) used in the second stage of estimation to recover the preference parameter \(\theta\). Each moment is defined based on observed application behavior, including indicators for listing specific centers in the initial or second ROL (\(R^1_i\) and \(R^2_i\)), interactions with initial priority scores (\(s_i^1\)), waitlist status in periods 1 and 2, dropping a safety option (\(DropSafety_i\)), and not reapplying in period 2. These moments are simulated for each applicant based on flow utilities and optimal ROLs, with 100 simulation draws used to calculate the targeted sample moments.
    
\end{tablenotes}
\end{threeparttable}
\caption{List of individual moments}
\label{tab:moments}
\end{table}

In computing these moments I need to be able to compute student $i$'s optimal pair of ROLs $R^1_i, R^2_i$ given a draw of $v_i$, given the first stage estimates of the assignment lotteries. 
For a given draw of $v_i$, a direct way of finding the optimal pair of ROLs is to search over every possible pair of ROLs. However, this is practically infeasible, since with $J=126$ schools and maximal list length of $K=5$, there are $|\mathcal{R}| \simeq 2.95 \times 10^{10}$ possible ROLs. Appendix \ref{sec:comp} introduces a greedy algorithm to compute the pair of ROLs that approximates the optimal pair of ROLs.

Denoting the vector of moments as $h_i(\theta)$, I minimize the following objective function:
\begin{equation}
    Q(\theta; \hat{S}^{-1}) = \left[\frac{1}{n}\sum_{i=1}^{n}{h_i(\theta)} \right]' \hat{S}^{-1} \left[\frac{1}{n}\sum_{i=1}^{n}{h_i(\theta)}\right]
\end{equation}
where
\begin{equation}
    \hat{S} = \frac{1}{n}\sum_{i=1}^{n}{(h_i(\theta_{init}) + u_i) (h_i(\theta_{init}) + u_i)'}
\end{equation}
Here, $u_i \sim N(0, I)$ is a noise independent of $h_i$ I add to ensure that $\hat{S}$ is invertible, and the initial value $\theta_{init}$ is chosen as:
\begin{equation}
    \theta_{init} = \argmin_{\theta}{Q(\theta; I)}
\end{equation}

\section{Estimation Results}

In this section I report the parameter estimates following the two step estimation strategy. The estimation is restricted to three cohorts: children who were age 0 in 2019, age 0 in 2020, and age 0 in 2021. This sample restriction is imposed because, for other cohorts, it is impossible to distinguish whether an applicant is applying for the first time or is already on the waitlist. 

\begin{figure}[!ht]
    \centering
    \begin{subfigure}[b]{0.48\linewidth}
        \centering
        \includegraphics[width=\linewidth]{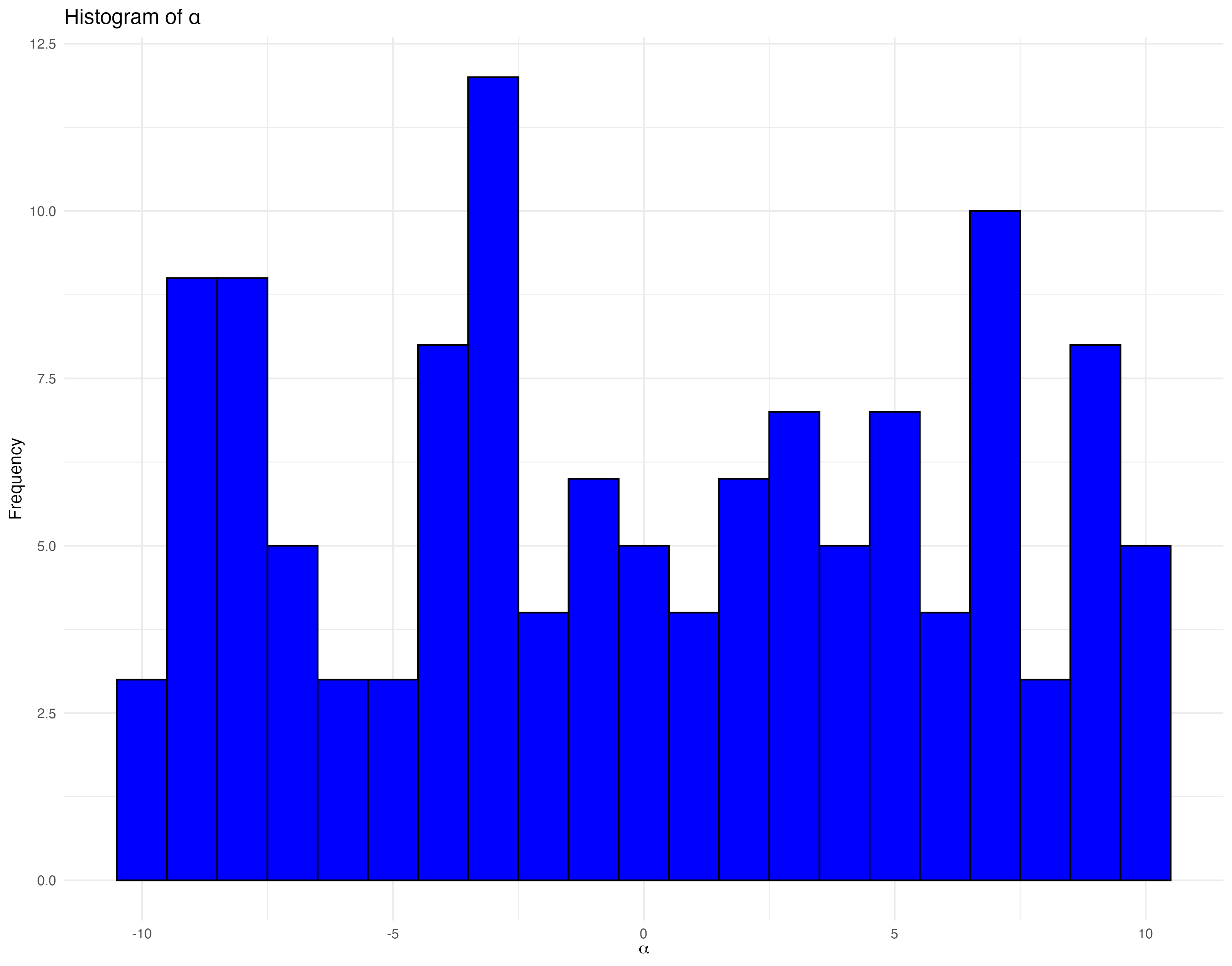}
        \caption{Histogram of $\hat{\alpha}$}
        \label{fig:alphahist}
    \end{subfigure}
    \hfill
    \begin{subfigure}[b]{0.48\linewidth}
        \centering
        \includegraphics[width=\linewidth]{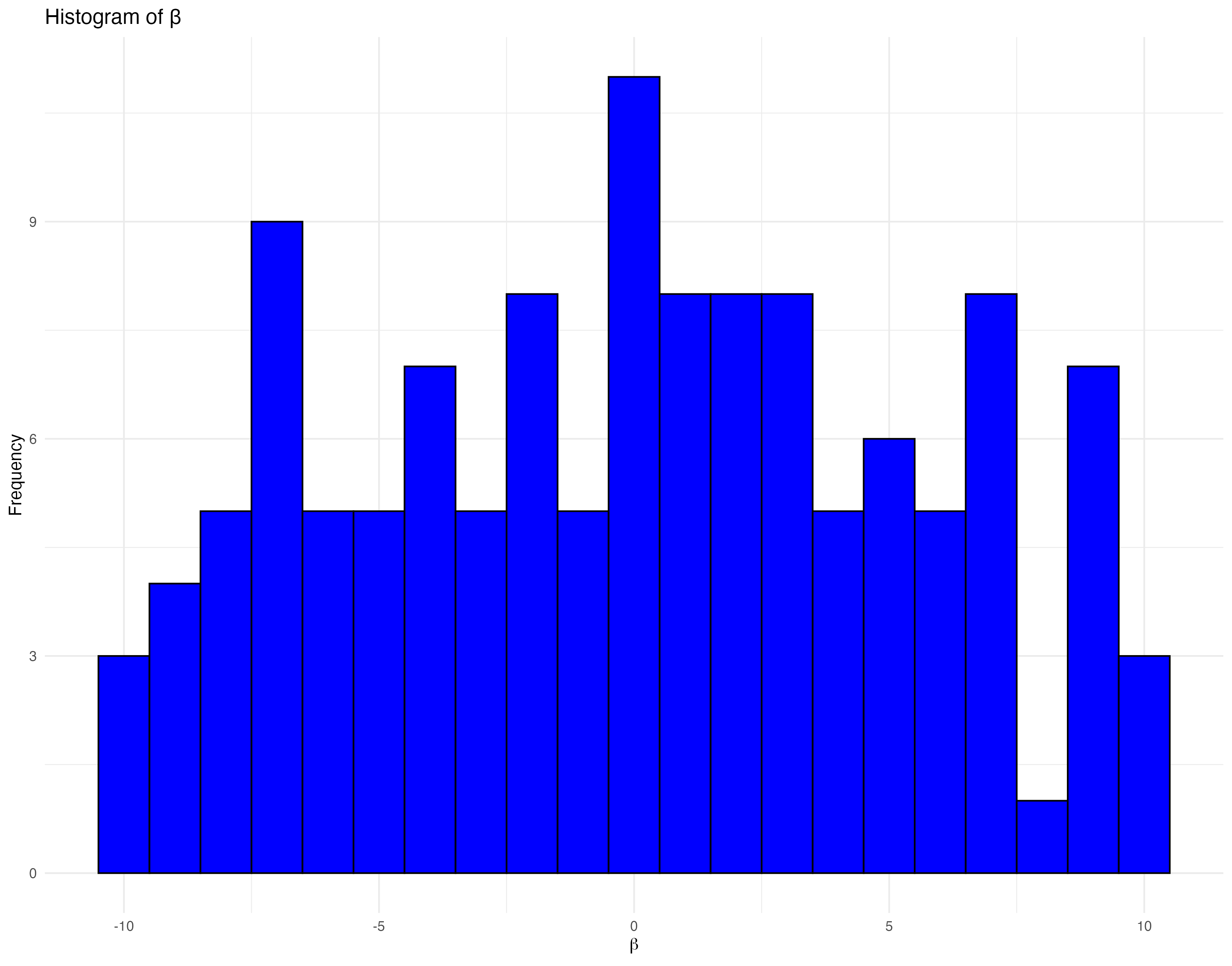}
        \caption{Histogram of $\hat{\beta}$}
        \label{fig:betahist}
    \end{subfigure}
    \caption{Histograms for $\hat{\alpha}$ and $\hat{\beta}$}
    \label{fig:alphabeta_hists}
    \begin{minipage}{0.75\textwidth} 
        {\scriptsize \textit{Note:} The histograms show the distribution of the point estimates for the daycare fixed effects, \(\hat{\alpha}_j\) (a; left), and the interaction terms between these fixed effects and the applicant's initial priority scores, \(\hat{\beta}_j\) (b; right), for each \(j \in \mathcal{J}\). The estimates are obtained from the second-stage preference parameter estimation.
        \par}
    \end{minipage}
\end{figure}

Figures \ref{fig:alphahist} and \ref{fig:betahist} present histograms of the point estimates for daycare fixed effects, $\alpha$, and their interaction with the applicant's initial priority score, $\beta_j$, for each $j \in \mathcal{J}$. Both estimates exhibit significant dispersion: the $\alpha_j$ estimates have a mean of 0.012 and a standard deviation of 6.038, while the $\beta_j$ estimates have a mean of -0.050 and a standard deviation of 5.476. 
To further investigate the factors influencing the vertical quality of daycare centers, I regress the estimates of $\hat{\alpha}_j$ and $\hat{\beta}_j$ on various daycare characteristics. These include the distance to the closest station (in kilometers), fixed effects for the nearest station, the total number of full-time staff, total capacity, an indicator for whether the center provides temporary childcare services, an indicator for whether the center has undergone a third-party review, and an indicator for whether it serves educational purposes as a certified facility combining daycare and education.
The OLS estimates are presented in Table \ref{tab:betaregression}, excluding the fixed effects for the nearest station. The first column corresponds to the specification using the daycare fixed effects, $\hat{\alpha}_j$, as the dependent variable. The results align with expectations: applicants prefer daycares that are closer to the station, have more staff, provide temporary childcare services, and have undergone a third-party review. Interestingly, the coefficient on total capacity is negative and significant, indicating a preference for smaller daycare centers over larger ones. The coefficient for educational purposes is positive but not statistically significant.
The second column uses the interaction with the applicant's priority score, $\hat{\beta}_j$, as the dependent variable. Here, the estimates generally have opposite signs, with significant results for distance and the total number of staff. This suggests that applicants with higher priority scores are less sensitive to long distances and fewer staff, likely due to a greater necessity for childcare services, which compels them to be less selective.

\begin{table}[htbp!] \centering 
  \caption{Regression Results} 
  \label{tab:betaregression} 

\begin{threeparttable}
\begin{tabular}{@{\extracolsep{5pt}}lcc} 
\hline \\[-1.8ex] 
 & \multicolumn{2}{c}{\textit{Dependent variable:}} \\ 
\cline{2-3} 
\\[-1.8ex] & $\hat{\alpha}_j$ & $\hat{\beta}_j$ \\ 
\\[-1.8ex] & (1) & (2)\\ 
\hline \\[-1.8ex] 
 Distance from the Closest Station & $-$0.485 & 0.410 \\ 
  & (0.210) & (0.193) \\ 
 Total Number of Full-Time Staff & 0.488 & $-$0.383 \\ 
  & (0.238) & (0.218) \\ 
 Total Capacity & $-$0.106 & 0.036 \\ 
  & (0.042) & (0.038) \\ 
 Temporary Childcare Service & 4.362 & $-$1.219 \\ 
  & (2.032) & (1.860) \\ 
 Third Party Review & 2.171 & $-$0.050 \\ 
  & (1.282) & (1.174) \\ 
 Education and Care Preschool & 11.305 & $-$3.626 \\ 
  & (6.845) & (6.267) \\ 
\hline \\[-1.8ex] 
Observations & 114 & 114 \\ 
R$^{2}$ & 0.321 & 0.286 \\ 
Adjusted R$^{2}$ & 0.127 & 0.083 \\ 
Residual Std. Error (df = 88) & 5.727 & 5.244 \\ 
F Statistic (df = 25; 88) & 1.660 & 1.408 \\ 
\hline \\[-1.8ex] 
\end{tabular} 

\begin{tablenotes}
    \scriptsize
    \item \textit{Note:}
    The dependent variables are $\hat{\alpha}_j$ (Column 1) and $\hat{\beta}_j$ (Column 2), obtained from the second stage of the structural estimation. 
    "Distance from the Closest Station" is measured in kilometers. 
    "Total Number of Full-Time Staff" refers to the total number of full-time employees. 
    "Total Capacity" indicates the total number of children the center can accommodate. 
    "Temporary Childcare Service" is an indicator variable equal to 1 if the center provides short-term childcare services. 
    "Third Party Review" is an indicator variable equal to 1 if the center has undergone an external evaluation. 
    "Education and Care Preschool" is an indicator variable equal to 1 if the center is certified as combining daycare and educational functions. 
    Fixed effects for the nearest station are included but not displayed. 
    Standard errors are reported in parentheses.
\end{tablenotes}

\end{threeparttable}

\end{table} 

Table \ref{tab:area_cov_matrix} presents the estimated variance matrix, $\Sigma$, of the unobserved heterogeneity, $\epsilon_{ij}$. Given that correlations are allowed only at the area level, I report the reduced 7×7 matrix at the area level instead of the full 126×126 matrix, $\hat{\Sigma}$.
\begin{table}[htbp!]
\centering
\begin{threeparttable}
\begin{tabular}{rrrrrrrr}
  \hline
 & A & B & C & D & E & F & G \\ 
  \hline
A & 52.32 & 94.94 & -144.64 & -42.40 & 93.63 & -102.14 & 49.94 \\ 
  B & 94.94 & 646.40 & -765.76 & -109.71 & -195.25 & 19.58 & 61.04 \\ 
  C & -144.64 & -765.76 & 1974.81 & 40.91 & 486.16 & 284.08 & -155.58 \\ 
  D & -42.40 & -109.71 & 40.91 & 58.32 & -38.26 & 87.73 & -27.83 \\ 
  E & 93.63 & -195.25 & 486.16 & -38.26 & 1343.68 & -80.84 & 128.75 \\ 
  F & -102.14 & 19.58 & 284.08 & 87.73 & -80.84 & 896.82 & -158.51 \\ 
  G & 49.94 & 61.04 & -155.58 & -27.83 & 128.75 & -158.51 & 70.45 \\ 
   \hline
\end{tabular}
\begin{tablenotes}
    \scriptsize \item \textit{Note:} This table presents the estimated variance matrix, $\Sigma$, of the unobserved heterogeneity, $\epsilon_{ij}$. Given that correlations are allowed only at the area level, I report the reduced 7×7 matrix at the area level instead of the full 126×126 matrix, $\hat{\Sigma}$.
\end{tablenotes}    
\end{threeparttable}

\caption{Area-Level Covariance Matrix} 
\label{tab:area_cov_matrix}
\end{table}

Table \ref{tab:outsideoption} presents the estimated preference parameters for the outside option, $\mu_0^a$ and $\sigma_0^{a2}$. Note that $\mu_0^0$ is normalized to zero for location invariance. As expected, the mean value of the outside option decreases from ages 0 to 1 and from 1 to 2. Interestingly, it also declines from age 2 to 3, which is counterintuitive given that applicants can begin utilizing kindergartens from age 3, potentially increasing the value of the outside option. However, this decline is accompanied by a substantial increase in variance.
One possible explanation for this pattern is that some small-scale daycare centers only accept children up to age 3. This limitation may reduce the relative value of the outside option for families transitioning from these facilities, resulting in lower mean values and higher dispersion.

\begin{table}[htbp!] 
\centering 
\caption{Estimates of Means and Variances of Outside Options} 
\label{tab:outside_estimates} 

\begin{threeparttable}
\begin{tabular}{ccc} 
\hline 
Age & Mean ($\mu_0$) & Variance ($\sigma_0^2$) \\
\hline 
0 & 0.000 & 25.432 \\
1 & -5.234 & 7.584 \\
2 & -6.634 & 15.709 \\
3 & -6.865 & 48.136 \\
4 & 1.457 & 57.988 \\
5 & -0.504 & 13.877 \\
\hline 
\end{tabular} 
\begin{tablenotes}
    \scriptsize \item \textit{Note:} The table reports the estimated preference parameters for the outside option, \(\mu_0^a\) (mean) and \(\sigma_0^{a2}\) (variance), for each age \(a\). The mean for age 0 (\(\mu_0^0\)) is normalized to zero for location invariance. 
\end{tablenotes}    
\end{threeparttable}

\label{tab:outsideoption}
\end{table}


\section{Counterfactual Analysis}

To understand the welfare implications of waitlist priority, I simulate equilibrium outcomes for different values of the additional priority \( b \) added to an applicant's initial priority score. The parameter \( b \) can take values from \(\{-1, 0, 1, 2, 3\}\), where \( b = 2 \) represents the baseline scenario, and \( b = 0 \) corresponds to the abolition of waitlist priority.

Using the second-stage estimates of preference parameters \(\hat{\theta}\), I simulate each applicant's flow utility for daycare centers (\(v_{ij}^m\)) and the outside option (\(v_{i0}^{am}\)) across simulation draws \(m = 1, 2, \dots, M\), with \( M = 7 \). Based on these flow utilities, I simulate the new equilibrium for each scenario \( b \in \{-1, 0, 1, 2, 3\} \). 

The rational expectations framework assumes that policy-induced changes in application behavior also affect applicants' beliefs about cutoffs. To incorporate this, I use an iterative procedure starting from the first-stage estimates of assignment probabilities, \(\hat{\Pi} = \{\hat{\pi}^t_j\}_{j = 1, \dots, J}^{t = 2019, \dots, 2021}\). For each counterfactual waitlist priority \( b \), the procedure updates applicants' beliefs and simulates assignments until convergence. The steps are as follows:

\begin{description}
    \item[] Step 0: Initialization 
    Set the initial belief about assignment probabilities:
    \[
    \Pi^0 = \hat{\Pi} = \{\hat{\pi}^t_j\}_{j = 1, \dots, J}^{t = 2019, \dots, 2021}.
    \]

    \item[] Step \( k \): Update Beliefs 
    Using the belief from step \( k-1 \), \(\Pi^{k-1}\), compute the optimal pair of ROLs, \((R^1_i, R^2_i)\), for each applicant \( i = 1, \dots, I \). Simulate assignments using the generated ROLs. For each bootstrap sample, compute the cutoffs for each center \( j \) and aggregate them to update the belief \(\Pi^k\).

\end{description}
I repeat step \( k \) until it meets the following convergence criterion:
    \[
    \frac{\lVert \Pi^k - \Pi^{k-1} \rVert}{\lVert \Pi^{k-1} \rVert} \leq \epsilon.
    \]
In each simulated equilibrium, indexed by \( b \) and \( m \), I compute the realized utilities \( V^{t}_i(b,m) \) for each applicant \( i \) in each period \( t \), based on their realized assignment. Additionally, I calculate the discounted sum of flow utilities as 
\[
V_i(b, m) = V^1_i(b,m) + \delta V^2_i(b,m).
\] 
These values are then used to compare welfare across different scenarios \( b \).

\begin{figure}
    \centering
    \includegraphics[width=0.9\linewidth]{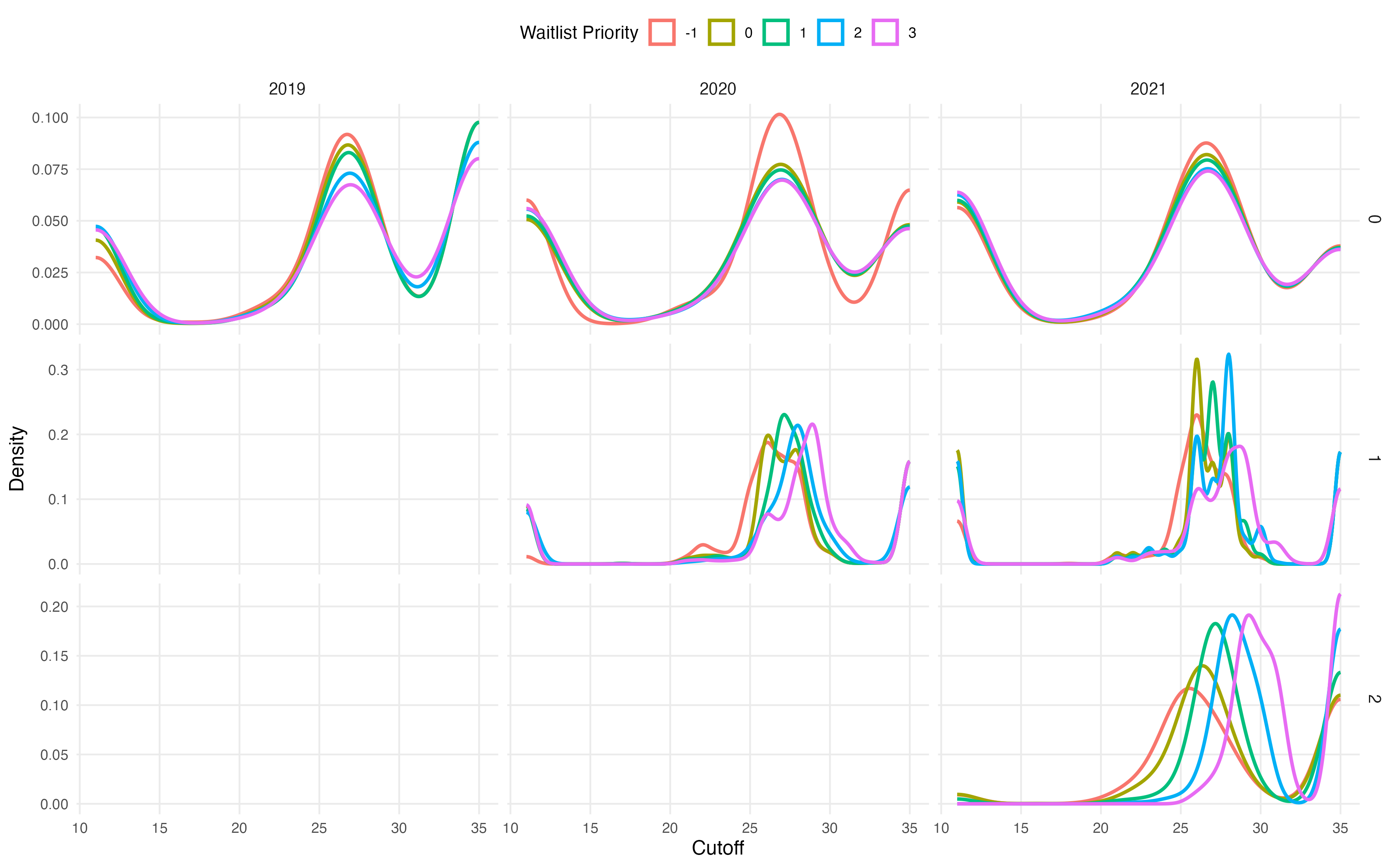}
    \caption{Distribution of Simulated Cutoffs by Waitlist Priority}
    \label{fig:counter_cutoffs}
    \begin{minipage}{0.75\textwidth} 
    {\scriptsize 
    \textit{Note:} The figure shows the density plots of simulated cutoff scores for daycare centers under different scenarios. Each column represents a year, and each row corresponds to an age group. Daycare centers with no vacancies are assigned a cutoff value of 35, while those without cutoffs (where the number of vacancies exceeds the number of applicants) are assigned a value of 11.
    \par}
\end{minipage}
\end{figure}

Figure \ref{fig:counter_cutoffs} illustrates the distribution of simulated cutoffs under each scenario. Each column of panels corresponds to a year, and each row to an age group. Daycare centers with no vacancies are assigned a cutoff value of 35, while those without cutoffs—where the number of vacancies exceeds the number of applicants—are assigned a value of 11. The impact of varying waitlist priorities differs across age groups.
For age 0, a decrease in waitlist priority primarily results in more daycare centers having cutoffs, without any significant shift in the peak of the distribution. Instead, the distribution becomes more concentrated around a cutoff value of 26. In contrast, for age 1, the peak of the distribution shifts leftward as waitlist priority decreases, indicating that previous waitlist priorities inflated the cutoffs for this age group. This leftward shift becomes even more pronounced for age 2, where the effect of reducing waitlist priority is most evident.

\begin{figure}
    \centering
    \includegraphics[width=0.9\linewidth]{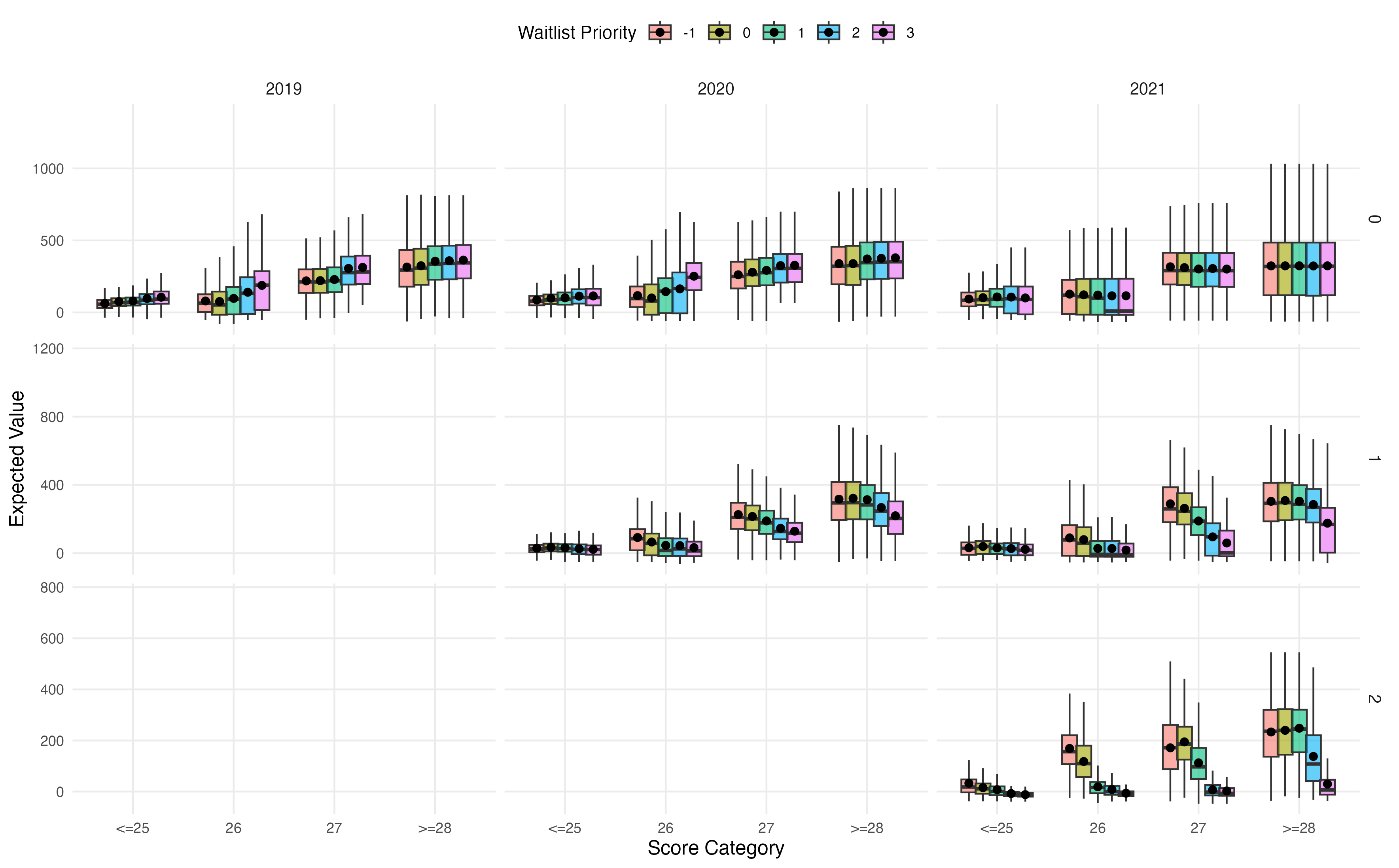}
    \caption{Utility Distributions by Waitlist Priority and Initial Score}
    \label{fig:counter_expv}
    \begin{minipage}{0.75\textwidth} 
        {\scriptsize Each panel shows boxplots of \(V_i(b,m)\), the utility of applicants, across scenarios \(b\), priority scores \(m\), and individuals \(i\). Columns represent different years, while rows correspond to the age of the applicant during their initial round of application. Priority scores are grouped into four categories: \(\leq 25\), \(26\), \(27\), and \(\geq 28\).

        \par}
    \end{minipage}
\end{figure}

Figure \ref{fig:counter_expv} shows boxplots of $V_i(b,m)$ across $m$ and $i$ for each scenario $b$ and the applicant't initial priority score. Each column of panels corresponds to a year, and each row to an age, corresponding to the applicant's initial round of application. The initial priority score is divided into four categories: $\leq 25, 26, 27$, and $\geq 28$. First thing to note is that applicants with higher initial scores receive higher utility in every case. This to some extent justifies the claim that the priority system is meant to capture the true beneficiaries of daycare service. 
The figure also highlights that the welfare implications of waitlist priority vary across different scores and age groups. Specifically, applicants whose initial age is 0 benefit from waitlist priority, particularly those with lower initial priority scores. In contrast, applicants whose initial age is 1 tend to lose out from waitlist priority, especially those with higher initial priority scores.

Let us examine the changes in applicants' application behavior in more detail. Table \ref{tab:simres} summarizes the mean list length, fraction of applicants waitlisted, mean $V^t$ in each period ($t=1,2$), and the discounted sum of utility $V$ across applicants by year, age, and scenario. 
Focusing on age 0 applicants, the list length in period 1 shortens as waitlist priority increases, reflecting stronger incentives for strategic waiting. This leads to a higher fraction of applicants being waitlisted in period 1 and a corresponding decrease in $V^1$. However, this trade-off results in higher utility in period 2, as $V^2$ increases with waitlist priority. Overall, the total utility $V$ improves for these applicants.
By contrast, age 1 applicants experience a sharp decline in $V^1$ as waitlist priority increases. This is primarily because they face intensified competition from applicants already on the waitlist. This negative effect carries over to $V^2$, as these applicants are less frequently assigned to their preferred daycares in period 1. As a result, the total utility $V$ decreases.
In summary, what initially appears to be a remedial measure for waitlisted applicants functions as a redistributive mechanism, shifting opportunities from late-starting, needy applicants to less-needy early starters.

\begin{table}[htbp!]
\centering\fontsize{10}{12}\selectfont
\resizebox{\linewidth}{!}{ 

\begin{threeparttable}
\begin{tabular}[t]{cccccccccc}
\toprule
\multicolumn{3}{c}{ } & \multicolumn{3}{c}{Period 1} & \multicolumn{3}{c}{Period 2} & \multicolumn{1}{c}{ } \\
\cmidrule(l{3pt}r{3pt}){4-6} \cmidrule(l{3pt}r{3pt}){7-9}
Year & Age & Waitlist Priority & List Length & Waitlisted & $V^1$ & List Length & Waitlisted & $V^2$ & $V$\\
\midrule
 &  & -1 & 4.03 & 0.36 & 24.11 & 4.78 & 0.17 & 26.22 & 156.70\\
\cmidrule{3-10}
 &  & 0 & 3.66 & 0.40 & 24.00 & 4.51 & 0.26 & 26.54 & 158.96\\
\cmidrule{3-10}
 &  & 1 & 3.36 & 0.44 & 23.86 & 4.28 & 0.21 & 30.79 & 179.58\\
\cmidrule{3-10}
 &  & 2 & 3.00 & 0.46 & 23.67 & 4.03 & 0.19 & 37.58 & 213.16\\
\cmidrule{3-10}
\multirow{-5}{*}{\centering\arraybackslash 2019} &  & 3 & 2.92 & 0.48 & 23.41 & 3.93 & 0.13 & 43.06 & 239.77\\
\cmidrule{1-1}
\cmidrule{3-10}
 &  & -1 & 4.02 & 0.29 & 30.99 & 4.70 & 0.11 & 33.99 & 201.85\\
\cmidrule{3-10}
 &  & 0 & 3.61 & 0.34 & 30.64 & 4.26 & 0.22 & 33.11 & 198.01\\
\cmidrule{3-10}
 &  & 1 & 3.44 & 0.36 & 30.53 & 4.04 & 0.15 & 39.67 & 229.93\\
\cmidrule{3-10}
 &  & 2 & 3.23 & 0.40 & 30.08 & 3.77 & 0.14 & 42.79 & 244.98\\
\cmidrule{3-10}
 & \multirow{-10}{*}{\centering\arraybackslash 0} & 3 & 3.19 & 0.40 & 29.91 & 3.74 & 0.06 & 51.04 & 285.49\\
\cmidrule{2-10}
 &  & -1 & 3.89 & 0.29 & 31.77 & 4.34 & 0.12 & 36.57 & 179.88\\
\cmidrule{3-10}
 &  & 0 & 3.81 & 0.33 & 29.52 & 4.08 & 0.17 & 34.61 & 170.57\\
\cmidrule{3-10}
 &  & 1 & 3.23 & 0.42 & 23.86 & 4.11 & 0.22 & 32.19 & 155.74\\
\cmidrule{3-10}
 &  & 2 & 3.11 & 0.52 & 15.70 & 4.22 & 0.24 & 28.28 & 132.25\\
\cmidrule{3-10}
\multirow{-10}{*}[0.5\dimexpr\aboverulesep+\belowrulesep+\cmidrulewidth]{\centering\arraybackslash 2020} & \multirow{-5}{*}{\centering\arraybackslash 1} & 3 & 2.74 & 0.60 & 9.02 & 4.26 & 0.28 & 23.93 & 108.55\\
\bottomrule
\end{tabular}

\begin{tablenotes}
    \scriptsize \item \textit{Note:} The table summarizes applicants' application behavior and utility outcomes by year, age, and waitlist priority scenarios. Period 1 and Period 2 metrics include mean list length, fraction of applicants waitlisted, and mean utility (\(V^1\) and \(V^2\), respectively). The total utility (\(V\)) represents the discounted sum of utility across both periods. Waitlist priority scenarios range from -1 to 3, with higher values indicating greater priority. 
\end{tablenotes}
    
\end{threeparttable}

}
\caption{Summary of Application Behavior and Utility Outcomes}
\label{tab:simres}
\end{table}

\section{Conclusion}
I focused on the Japanese daycare system and demonstrated the prevalence of manipulation arising from the additional priority granted to waitlisted applicants. I showed that many waitlisted applicants avoid listing safety options in their initial applications, only to include them when reapplying. This behavior is driven by the potential benefits of being waitlisted, specifically the increased probability of admission to more selective daycare centers. These findings support the claim that applicants respond to dynamic incentives, providing evidence against alternative explanations such as declining outside option values or updated preferences.

The counterfactual exercise, based on structural estimates, reveals that removing additional priority for waitlisted applicants reduces competition and encourages more applicants to use accredited daycare centers at earlier ages, instead of strategically waiting. The primary welfare effect is redistributive: it diminishes the advantage of early starters and reallocates opportunities to later applicants who may have greater immediate need. These findings highlight the possible unintended consequences of well-meaning policies and emphasize the importance of designing assignment mechanisms that promote fairness and efficiency without encouraging strategic manipulation.

This analysis has several limitations. First, I model market entry as exogenous, overlooking how parents might strategically time their initial applications based on the duration of their parental leave. Second, while I account for variation in the value of outside options, I assume stable preferences for accredited daycare over time, excluding the possibility of learning or other forms of preference evolution. As a result, the flow utility should be interpreted as an ex-ante expected value. Third, and most importantly, this study focuses on dynamic incentives arising in markets that allocate goods simultaneously but also allow for reapplication, granting explicit priority to waitlisted participants. While this feature might be unique to the Japanese daycare system, similar dynamics can emerge in school choice mechanisms, such as the Delaware school choice example discussed in the introduction, where waitlisting also incentivizes strategic waiting among applicants. This paper highlights the risks of prioritizing waitlisted participants, cautioning that such policies can lead to unintended consequences.


\clearpage

\section*{Appendix}
\section*{Details of the Priority Score}
\begin{table}[htbp!]
    \centering
    \resizebox{\textwidth}{!}{%
    \begin{threeparttable}
    \begin{tabular}{|l|l|l|l|}
    \hline
    \multicolumn{2}{|c|}{\begin{tabular}[c]{@{}c@{}}Basic   score\\      (each for the mother and the father)\end{tabular}}                                                                                                                                                                                                                                                                                                                                                                                                                                                                                                    & \multicolumn{2}{c|}{Adjustment score}                                                                                               \\ \hline
    \multicolumn{1}{|c|}{Item}                                                                                                                                                                                                                                                                                                                        & \multicolumn{1}{c|}{Point}                                                                                                                                                                                                                                             & \multicolumn{1}{c|}{Item}                                                                              & \multicolumn{1}{c|}{Point} \\ \hline
    Is   employed                                                                                                                                                                                                                                                                                                                                     & 5 - 10                                                                                                                                                                                                                                 & Is a resident of the   municipality                                                                    & 1   - 4   \\
    Is   seeking for a job                                                                                                                                                                                                                                                                                                                            & 5                                                                                                                                                                                                                                                                      & \begin{tabular}[c]{@{}l@{}} (*) Is not an incumbent user of \\ a accredited daycare center\end{tabular} & 1                          \\
    Is a   student                                                                                                                                                                                                                                                                                                                                    & 6 - 8                                                                                                                                                                                                                                                                  & Is on welfare                                                                                          & 4                          \\
    Is giving   a birth                                                                                                                                                                                                                                                                                                                                & 7                                                                                                                                                                                                                                                                      & The child is single parented                                                                           & 1 - 3                      \\
    Is ill /   has a disability                                                                                                                                                                                                                                                                                                                       & 6 - 10                                                                                                                                                                                                                                                                 & The child has siblings                                                                                 & 1 - 2                      \\
    Needs   nursing                                                                                                                                                                                                                                                                                                                                   & 6 - 10                                                                                                                                                                                                                                                                 & The child is disabled                                                                                  & 1 - 2                      \\
    \begin{tabular}[c]{@{}l@{}}Can't take care of the child \\ because of a natural disaster\end{tabular}                                                                                                                                                                                                                                             & 10                                                                                                                                                                                                                                                                     & \begin{tabular}[c]{@{}l@{}}(*) Is using a non-accredited \\ daycare center\end{tabular}                & 1                          \\
    Doesn't   exist                                                                                                                                                                                                                                                                                                                                   & 10                                                                                                                                                                                                                                                                     & (*) Is wait-listed                                                                                      & 1                          \\ \cline{1-2}
    \multicolumn{2}{|c|}{Tie breakers}                                                                                                                                                                                                                                                                                                                                                                                                                                                                                                                                                                                         & \begin{tabular}[c]{@{}l@{}}Does not have grandparents \\ nearby to look after the child\end{tabular}   & 1                          \\ \cline{1-2}
    \multicolumn{2}{|l|}{\multirow{7}{*}{\begin{tabular}[c]{@{}l@{}}Resident \textgreater~ No arrears \textgreater~ On welfare \textgreater~ \\ Single parent \textgreater~ No defers \textgreater~ Disabilities \textgreater~\\ Basic scores \textgreater~ After paternal leave \textgreater~\\ Is graduating~ \textgreater~ Is a childcare worker \textgreater~ \\ Has siblings \textgreater~ Parent type \textgreater~ Not an incumbent~ \textgreater~\\ Parents are separated \textgreater~ Multiple birth \textgreater~ \\ Using a non-accredited center~ \textgreater~ Is wait-listed~ \textgreater~ Lower income \textgreater~\\ Time period as a resident\end{tabular}}} & \begin{tabular}[c]{@{}l@{}}Has received an offer of \\ employment\end{tabular}                         & 1                          \\
    \multicolumn{2}{|l|}{}                                                                                                                                                                                                                                                                                                                                                                                                                                                                                                                                                                                                     & Is graduating after age 2 or 3                                                                         & 2                          \\
    \multicolumn{2}{|l|}{}                                                                                                                                                                                                                                                                                                                                                                                                                                                                                                                                                                                                     & \begin{tabular}[c]{@{}l@{}}Is helping a family operated \\ business\end{tabular}                       & -1                         \\
    \multicolumn{2}{|l|}{}                                                                                                                                                                                                                                                                                                                                                                                                                                                                                                                                                                                                     & Has become unemployed                                                                                  & 2                          \\
    \multicolumn{2}{|l|}{}                                                                                                                                                                                                                                                                                                                                                                                                                                                                                                                                                                                                     & Has returned from parental leave                                                                       & 3                          \\
    \multicolumn{2}{|l|}{}                                                                                                                                                                                                                                                                                                                                                                                                                                                                                                                                                                                                     & Has deferred                                                                                           & No points added for (*)    \\
    \multicolumn{2}{|l|}{}                                                                                                                                                                                                                                                                                                                                                                                                                                                                                                                                                                                                     & \begin{tabular}[c]{@{}l@{}}Is in arrears of the childcare \\ fee\end{tabular}                          & No adjustment scores added \\ \hline
    \end{tabular}%
    
    \begin{tablenotes}
        \scriptsize
        \item \textit{Note:} Translated by the author.
    \end{tablenotes}    
    \end{threeparttable}
    
    }
    \caption{Decomposition of Priority Score}
    \label{tab:priorityscore}
\end{table}

\section*{Summary Statistics}

\begin{table}[htbp!]
\centering
\caption{Summary Statistics for Daycare Centers (Full Sample)}
\begin{tabular}{lcccccc}
  \hline
 & Age 0 & Age 1 & Age 2 & Age 3 & Age 4 & Age 5 \\ 
  \hline
  \textit{Year 2019} \\
Number of centers & 79 & 93 & 93 & 70 & 61 & 61 \\ 
  Number of public centers & 16 & 21 & 21 & 20 & 17 & 17 \\ 
  Total capacity & 577 & 984 & 1100 & 1056 & 1952 & 1952 \\ 
  Mean capacity & 7.30 & 10.58 & 11.83 & 15.09 & 32.00 & 32.00 \\ 
  Number of centers with vacant seats & 78 & 83 & 67 & 44 & 31 & 23 \\ 
  Total vacant seats & 560 & 515 & 261 & 261 & 202 & 80 \\ 
  Mean vacant seats & 7.21 & 6.20 & 3.90 & 6.02 & 6.65 & 3.48 \\ 
  Mean total applications & 38.05 & 36.37 & 20.75 & 20.16 & 13.29 & 2.30 \\ 
  Mean top ranked applications & 9.89 & 9.01 & 5.89 & 5.33 & 3.13 & 0.61 \\ 
  Fraction of no cutoffs & 0.11 & 0.05 & 0.06 & 0.40 & 0.52 & 0.74 \\ 
  Mean cutoff score & 26.04 & 26.96 & 26.98 & 28.12 & 28.87 & 27.00 \\ 
   \hline
   \textit{Year 2020} \\
Number of centers & 95 & 112 & 112 & 86 & 77 & 77 \\ 
  Number of public centers & 16 & 21 & 21 & 20 & 17 & 17 \\ 
  Total capacity & 661 & 1137 & 1272 & 1263 & 2404 & 2404 \\ 
  Mean capacity & 6.96 & 10.15 & 11.36 & 14.69 & 31.22 & 31.22 \\ 
  Number of centers with vacant seats & 95 & 100 & 80 & 55 & 49 & 29 \\ 
  Total vacant seats & 645 & 574 & 286 & 284 & 293 & 181 \\ 
  Mean vacant seats & 6.81 & 5.75 & 3.57 & 5.24 & 6.10 & 6.24 \\ 
  Mean total applications & 28.42 & 32.25 & 12.46 & 17.71 & 6.22 & 2.00 \\ 
  Mean top ranked applications & 7.30 & 7.73 & 3.14 & 4.30 & 1.35 & 0.38 \\ 
  Fraction of no cutoffs & 0.33 & 0.03 & 0.33 & 0.36 & 0.71 & 0.82 \\ 
  Mean cutoff score & 25.74 & 26.43 & 26.81 & 27.34 & 27.50 & 26.80 \\ 
   \hline
  \textit{Year 2021} \\
Number of centers & 96 & 113 & 113 & 87 & 77 & 77 \\ 
  Number of public centers & 16 & 21 & 21 & 20 & 17 & 17 \\ 
  Total capacity & 667 & 1151 & 1286 & 1269 & 2404 & 2404 \\ 
  Mean capacity & 6.95 & 10.19 & 11.38 & 14.59 & 31.22 & 31.22 \\ 
  Number of centers with vacant seats & 102 & 107 & 83 & 69 & 54 & 52 \\ 
  Total vacant seats & 644 & 514 & 166 & 217 & 173 & 294 \\ 
  Mean vacant seats & 6.85 & 5.41 & 2.81 & 4.51 & 5.54 & 5.90 \\ 
  Mean total applications & 25.37 & 25.05 & 12.87 & 9.32 & 4.93 & 1.25 \\ 
  Mean top ranked applications & 6.47 & 5.95 & 3.19 & 2.25 & 1.13 & 0.27 \\ 
  Fraction of no cutoffs & 0.41 & 0.19 & 0.20 & 0.67 & 0.85 & 0.85 \\ 
  Mean cutoff score & 25.61 & 26.02 & 26.12 & 27.17 & 26.88 & 27.25 \\ 
   \hline
\end{tabular}
\label{sumstat_schools_full}
\end{table}

\begin{table}[htbp!]
\centering
\caption{Summary Statistics for Applicants (Full Sample)}
\resizebox{!}{0.5 \textheight}{
\begin{tabular}{lcccccc}
  \hline
 & Age 0 & Age 1 & Age 2 & Age 3 & Age 4 & Age 5 \\ 
  \hline
  \textit{Year 2019} \\
Number of applications & 732 & 762 & 424 & 282 & 165 & 36 \\ 
  Mean priority score & 26.37 & 26.70 & 26.61 & 27.15 & 27.22 & 26.44 \\ 
  Fraction of already waitlisted applicants & - & - & - & - & - & - \\ 
  Fraction of incumbent applicants & 0.000 & 0.039 & 0.120 & 0.167 & 0.230 & 0.250 \\ 
  Mean list length & 4.02 & 4.02 & 3.68 & 3.65 & 3.32 & 2.39 \\ 
  Fraction of top choice in the same area & 0.261 & 0.210 & 0.151 & 0.080 & 0.087 & 0.087 \\ 
  Fraction of being assigned to first choice & 0.475 & 0.383 & 0.297 & 0.461 & 0.333 & 0.333 \\ 
  Fraction of being assigned to second choice & 0.122 & 0.118 & 0.156 & 0.113 & 0.103 & 0.056 \\ 
  Fraction of being assigned to third choice & 0.068 & 0.076 & 0.094 & 0.060 & 0.115 & 0.056 \\ 
  Fraction of being assigned to fourth choice & 0.042 & 0.045 & 0.038 & 0.039 & 0.042 & 0.056 \\ 
  Fraction of being assigned to fifth choice & 0.026 & 0.031 & 0.028 & 0.007 & 0.024 & 0.000 \\ 
  Fraction of ending up unassigned & 0.266 & 0.318 & 0.283 & 0.149 & 0.200 & 0.333 \\ 
   \hline
\textit{Year 2020} \\
Number of applications & 660 & 774 & 275 & 290 & 122 & 40 \\ 
  Mean priority score & 26.45 & 26.64 & 26.36 & 26.95 & 27.34 & 27.25 \\ 
  Fraction of already waitlisted applicants & 0.000 & 0.169 & 0.342 & 0.121 & 0.115 & 0.300 \\ 
  Fraction of incumbent applicants & 0.002 & 0.047 & 0.196 & 0.217 & 0.246 & 0.200 \\ 
  Mean list length & 4.00 & 4.07 & 3.80 & 3.83 & 3.08 & 2.65 \\ 
  Fraction of top choice in the same area & 0.269 & 0.277 & 0.228 & 0.144 & 0.176 & 0.000 \\ 
  Fraction of being assigned to first choice & 0.582 & 0.416 & 0.385 & 0.390 & 0.533 & 0.350 \\ 
  Fraction of being assigned to second choice & 0.158 & 0.146 & 0.229 & 0.224 & 0.098 & 0.050 \\ 
  Fraction of being assigned to third choice & 0.068 & 0.083 & 0.076 & 0.076 & 0.041 & 0.075 \\ 
  Fraction of being assigned to fourth choice & 0.038 & 0.053 & 0.044 & 0.045 & 0.033 & 0.050 \\ 
  Fraction of being assigned to fifth choice & 0.012 & 0.031 & 0.022 & 0.010 & 0.008 & 0.025 \\ 
  Fraction of ending up unassigned & 0.142 & 0.247 & 0.171 & 0.110 & 0.107 & 0.350 \\ 
   \hline
   \textit{Year 2021} \\
Number of applications & 634 & 637 & 292 & 181 & 99 & 39 \\ 
  Mean priority score & 26.44 & 26.46 & 26.10 & 26.48 & 27.07 & 26.95 \\ 
  Fraction of already waitlisted applicants & 0.000 & 0.080 & 0.284 & 0.088 & 0.051 & 0.154 \\ 
  Fraction of incumbent applicants & 0.003 & 0.050 & 0.216 & 0.331 & 0.323 & 0.359 \\ 
  Mean list length & 3.93 & 4.06 & 3.74 & 3.65 & 3.13 & 2.54 \\ 
  Fraction of top choice in the same area & 0.348 & 0.308 & 0.273 & 0.159 & 0.190 & 0.000 \\ 
  Fraction of being assigned to first choice & 0.672 & 0.493 & 0.408 & 0.475 & 0.475 & 0.333 \\ 
  Fraction of being assigned to second choice & 0.136 & 0.190 & 0.192 & 0.254 & 0.121 & 0.051 \\ 
  Fraction of being assigned to third choice & 0.058 & 0.072 & 0.079 & 0.094 & 0.051 & 0.077 \\ 
  Fraction of being assigned to fourth choice & 0.024 & 0.039 & 0.051 & 0.017 & 0.010 & 0.000 \\ 
  Fraction of being assigned to fifth choice & 0.014 & 0.020 & 0.017 & 0.006 & 0.010 & 0.026 \\ 
  Fraction of ending up unassigned & 0.096 & 0.160 & 0.134 & 0.039 & 0.101 & 0.205 \\ 
   \hline
\end{tabular}
}
\label{sumstat_students_full}
\end{table}

\section*{Computation of the Optimal Pair of ROLs}\label{sec:comp}

For expositional purposes, in this section I will assume that the daycare centers listed in $R^1$ and $R^2$ are listed in descending order of the applicant's flow utility.
It is useful to first define the objective function:
\begin{equation}
    V(R^1, R^2) = 
    v \cdot \Tilde{L}(R^1, R^2) 
    + \left( v^{a_0}_{0} p^1(R^1) + \delta v^{a_0+1}_{0} p^2(R^2) + \dotsb + \delta^{5 - a_0} v^5_{0}p^2(R^2) \right)
\end{equation}
where I have dropped the subscript $i$ for notational convenience. Let $(R^{1*}, R^{2*})$ denote the optimal pair of ROLs:
\begin{equation}
    (R^{1*}, R^{2*}) = \argmax_{R^1 \in \mathcal{R}, R^2 \in \mathcal{R} \cup \{0\}\}}{V(R^1, R^2)}
\end{equation}

Also denote $\bar{R}^t$ the ROL that maximizes the flow expected utility in each period:
\begin{equation}
    \bar{R}^1 = \argmax_{R^1 \in \mathcal{R}}{V^1(R^1)}
\end{equation}
where
\begin{equation}\label{eq:val1}
    V^1(R^1) = v \cdot L^1(R^1) + v_0^1 p^1(R^1)
\end{equation}
and 
\begin{equation}
    \bar{R}^2 = \argmax_{R^2 \in \mathcal{R} \cup \{0\}}{V^2(R^2)}
\end{equation}
where
\begin{equation}\label{eq:val2}
\begin{aligned}
    V^2(R^2) &= {(v + \delta v + \dotsb + \delta^{5 - a_0} v) \cdot L^2(R^2) + (v_0^{a_0+1} + \delta v_0^{a_0+2} + \dotsb + \delta^{5-a_0} v_0^{5}) p^2(R^2)}\nonumber \\
    &= {v \cdot L^2(R^2) + \Tilde{v}_0^2 p^2(R^2)}  
\end{aligned}    
\end{equation}
where $\Tilde{v}_0^2 = \frac{v_0^{a_0+1} + \delta v_0^{a_0+2} + \dotsb + \delta^{5-a_0} v_0^{5}}{1 + \delta + \dotsb + \delta^{5-a_0}}$ is the value of outside option in period 2, normalized to per year value.

Both $\bar{R}^1$ and $\bar{R}^2$ can be computed using the marginal improvement algorithm by \cite{chadeSimultaneousSearch2006}, shown below in \ref{def:mia} since each maximand has their \textit{downward recursive structure}. To see this, note that the maximand of (\ref{eq:val1}) can be rewritten as follows:
\begin{equation}
    v \cdot L^1(R^1) + v_0^1 p^1(R^1) = \sum_{k=1}^{K+1} v_{R_k^1} \, \pi^1_{R^1_k} \prod_{k' < k} (1 - \pi^1_{R_{k'}})
\end{equation}
where $K = |R^1|$ and with an abuse of notation I define $R^1_{K+1}$ as the outside option, so that $v_{R^1_{K+1}} = v_0^{a_0+1}$ and $\pi^1_{R^1_{|K|+1}}=1$. The maximand of (\ref{eq:val2}) can be rewritten similarly.

\begin{definition}{Marginal Improvement Algorithm (MIA) (Chade and Smith, 2006)}\label{def:mia}
    \begin{itemize}
        \item[] {Step 0:} Let $R_0 = 0$.
        \item[] {Step 1:} Choose any $j_n \in \argmax\limits_{j \in \mathcal{J} \setminus R_{n-1}}
        {V^t(R_{n-1} \cup \{j\})}
        $.
        \item[] {Step 2:} If $V^t(R_{n-1} \cup \{j_n\}) < V^t(R_{n-1})$,
        then stop.
        \item[] {Step 3:} Set $R_n = R_{n-1} \cup \{j_n\}$ and go to Step 1.
    \end{itemize}    
\end{definition}
\noindent 
Note that I would only have to evaluate the objective function less than $K \times J$ times under the MIA instead of for every possible $|\mathcal{R}|$ ROLs.

I construct an approximation to $(R^{1*}, R^{2*})$ from $(\bar{R}^1), \bar{R}^2)$ as follows. First, $R^{2*} = \bar{R}^{2}$, since the choice of $R^2$ does not affect the expected flow utility in period 1. Given $\bar{R}^t$ ($t=1,2$), I develop an algorithm (\ref{def:approxalgo}) to approximate $R^{1*}$. This algorithm iteratively updates $\bar{R}^1$ either by dropping one listed school or swapping one listed school with another.
\begin{definition} {Approximation Algorithm} \label{def:approxalgo}
    \begin{itemize}
        \item[] {Step 0:} Let $R^{1\#}_0 = \bar{R}^1$.
        \item[] {Step 1:} Choose any $(k_n, j_n) \in \argmax\limits_{k \in R^{1\#}_{n-1}, j \in \{0\} \cup (\mathcal{J} \setminus R^{1\#}_{n-1})} 
        {V((R^1 \setminus \{k\}) \cup \{j\}), \bar{R}^2))}$
        \item[] {Step 2:} If 
        $V((R^1 \setminus \{k_n\}) \cup \{j_n\}), \bar{R}^2) < V(R^{1\#}_{n-1}, \bar{R}^2)$, then stop.
        \item[] {Step 3:} Set $R^{1\#}_{n} = 
        (R^1 \setminus \{k_n\}) \cup \{j_n\}
        $.
    \end{itemize}    
\end{definition}
\noindent Note that Step 2 requires me to evaluate the objective function less than $K \times J$ times. 

To see how well $R^{1\#}$ approximates $R^{1*}$, I conduct a simulation exercise with $K=3$ and $J=10$.
To do this, I draw 10 schools, and use the first stage estimates of the assignment probabilities described in Section \ref{sec:firststage}. To see how having higher utility for more selective schools induces deviation from $\bar{R}^{1}$, I simulate $M=1,000$ draws of $v_i$'s in the following way: first, independently for each school $j$, I obtain $M$ draws of ${v}_{ij} \sim \mathcal{N}(c (1 - \pi_j), 1)$, where $\pi_j$ is the first stage estimate of assignment probability for submitting a ROL consisting only of school $j$, for the year 2019, age 0, and priority score 26, and $c \in \{0.0, 0.5, 1.0\}$. For each simulated $v_i$, I compute $(R^{1\#}, R^{2\#})$ using the approximation algorithm and compare it to the directly computed $(R^{1*}, R^{2*})$.

As the first row of Table \ref{tab:algosim} shows, the approximation algorithm performs reasonably well in finding $R^{1*}$. The first row reports the fraction of simulation draws in which $V(R^{1\#}, R^{2\#}) \geq V(R^{1*}, R^{2*})$
The next 6 rows show how often $R^{1*}$ is updated as defined in Steps 1-3 of the approximation algorithm to construct $\bar{R}^1$. While for $c=0$, $R^{1*}$ does not differ from $\bar{R}^1$ 83.9\% of times, when $c$ is higher, deviation is larger and more frequent. 

\begin{table}[htbp!]
    \centering
    \begin{threeparttable}
    \begin{tabular}{lccc}
        \hline
        c & 0.0 & 1.0 & 2.0 \\
        \hline
        Correctly computed (\%) & 0.8 & 0.98 & 0.95 \\
        0 updates (\%) & 0.45 & 0.76 & 0.75 \\
        1 update (\%) & 0.22 & 0.21 & 0.21 \\
        2 updates (\%) & 0.16 & 0.01 & 0.02 \\
        3 updates (\%) & 0.15 & 0.02 & 0.02 \\
        4 updates (\%) & 0.01 & 0.0 & 0.0 \\
        More than 5 updates (\%) & 0.0 & 0.0 & 0.0 \\
        \hline
    \end{tabular}
    \begin{tablenotes}
        \scriptsize \item \textit{Note:} The table presents the results of a simulation exercise evaluating the performance of the approximation algorithm in computing the optimal rank-ordered list (ROL) \(R^{1*}\). The first row shows the percentage of simulation draws where the utility achieved by the approximated ROL \((R^{1\#}, R^{2\#})\) matches or exceeds that of the directly computed optimal ROL \((R^{1*}, R^{2*})\). The subsequent rows report the frequency of updates required to transition from \(\bar{R}^1\) to \(R^{1*}\) under different values of \(c\). The parameter \(c\) represents the utility disparity between selective and non-selective schools and is defined as follows: For each school \(j\), \(v_{ij}\) is drawn from \(\mathcal{N}(c (1 - \pi_j), 1)\), where \(\pi_j\) is the first stage estimate of the assignment probability for submitting a ROL consisting only of school \(j\) (for the year 2019, age 0, and priority score 26). Larger values of \(c\) increase the utility for more selective schools, leading to greater deviations from \(\bar{R}^1\).

    \end{tablenotes}        
    \end{threeparttable}
    \caption{Simulation Results for the Approximation Algorithm}
    \label{tab:algosim}
\end{table}




\printbibliography

@article{agarwalDemandAnalysisUsing2018,
  title = {Demand {{Analysis Using Strategic Reports}}: {{An Application}} to a {{School Choice Mechanism}}},
  shorttitle = {Demand {{Analysis Using Strategic Reports}}},
  author = {Agarwal, Nikhil and Somaini, Paulo},
  year = {2018},
  journal = {Econometrica},
  volume = {86},
  number = {2},
  pages = {391--444},
  issn = {1468-0262},
  doi = {10.3982/ECTA13615},
  urldate = {2023-07-10},
  copyright = {{\copyright} 2018 The Econometric Society},
  langid = {english}
}

@article{agarwalEquilibriumAllocationsAlternative,
  title = {Equilibrium {{Allocations Under Alternative Waitlist Designs}}: {{Evidence From Deceased Donor Kidneys}}},
  author = {Agarwal, Nikhil and Ashlagi, Itai and Rees, Michael A and Somaini, Paulo and Waldinger, Daniel},
  langid = {english},
  year = {2021},
  journal = {Econometrica}
}

@article{asaiChildcareAvailabilityHousehold2015,
  title = {Childcare Availability, Household Structure, and Maternal Employment},
  author = {Asai, Yukiko and Kambayashi, Ryo and Yamaguchi, Shintaro},
  year = {2015},
  month = feb,
  journal = {Journal of the Japanese and International Economies},
  volume = {38},
  pages = {172--192},
  doi = {10.2139/ssrn.2462366},
  langid = {english}
}

@article{calsamigliaStructuralEstimationModel2020,
  title = {Structural {{Estimation}} of a {{Model}} of {{School Choices}}: {{The Boston Mechanism}} versus {{Its Alternatives}}},
  shorttitle = {Structural {{Estimation}} of a {{Model}} of {{School Choices}}},
  author = {Calsamiglia, Caterina and Fu, Chao and G{\"u}ell, Maia},
  year = {2020},
  month = feb,
  journal = {Journal of Political Economy},
  volume = {128},
  number = {2},
  pages = {642--680},
  publisher = {The University of Chicago Press},
  issn = {0022-3808},
  doi = {10.1086/704573},
  urldate = {2023-07-10}
}

@article{chadeSimultaneousSearch2006,
  title = {Simultaneous {{Search}}},
  author = {Chade, Hector and Smith, Lones},
  year = {2006},
  journal = {Econometrica},
  volume = {74},
  number = {5},
  pages = {1293--1307},
  issn = {1468-0262},
  doi = {10.1111/j.1468-0262.2006.00705.x},
  urldate = {2023-07-10},
  langid = {english}
}

@article{fukaiChildcareAvailabilityFertility2017,
  title = {Childcare Availability and Fertility: {{Evidence}} from Municipalities in {{Japan}}},
  shorttitle = {Childcare Availability and Fertility},
  author = {Fukai, Taiyo},
  year = {2017},
  month = mar,
  journal = {Journal of the Japanese and International Economies},
  volume = {43},
  pages = {1--18},
  issn = {0889-1583},
  doi = {10.1016/j.jjie.2016.11.003},
  urldate = {2023-07-16},
  langid = {english}
}

@unpublished{hahmDynamicFrameworkSchool,
  title = {A Dynamic Framework of School Choice: Effects of Middle Schools on High School Choice},
  author = {Hahm, Dong Woo and Park, Minseon},
  langid = {english},
  year = {2022},
  note = {Working paper}
}

@article{kamadaFairMatchingConstraints2023,
  title = {Fair Matching under Constraints: Theory and Applications},
  shorttitle = {Fair Matching under Constraints},
  author = {Kamada, Yuichiro and Kojima, Fuhito},
  year = {2024},
  month = apr,
  journal = {The Review of Economic Studies},
  volume = {91},
  number = {2},
  pages = {1162--1199},
  issn = {0034-6527},
  doi = {10.1093/restud/rdad046},
  urldate = {2023-07-18}
}

@article{kaporHeterogeneousBeliefsSchool2020,
  title = {Heterogeneous {{Beliefs}} and {{School Choice Mechanisms}}},
  author = {Kapor, Adam J. and Neilson, Christopher A. and Zimmerman, Seth D.},
  year = {2020},
  month = may,
  journal = {American Economic Review},
  volume = {110},
  number = {5},
  pages = {1274--1315},
  issn = {0002-8282},
  doi = {10.1257/aer.20170129},
  urldate = {2024-03-29},
  langid = {english}
}

@article{kennesDayCareAssignment2014,
  title = {The {{Day Care Assignment}}: {{A Dynamic Matching Problem}}},
  shorttitle = {The {{Day Care Assignment}}},
  author = {Kennes, John and Monte, Daniel and Tumennasan, Norovsambuu},
  year = {2014},
  month = nov,
  journal = {American Economic Journal: Microeconomics},
  volume = {6},
  number = {4},
  pages = {362--406},
  issn = {1945-7669, 1945-7685},
  doi = {10.1257/mic.6.4.362},
  urldate = {2024-03-25},
  langid = {english}
}

@unpublished{larroucauDynamicCollegeAdmissions,
  title = {Dynamic College Admissions},
  author = {Larroucau, Tomas and Rios, Ignacio},
  year = {2022},
  note = {Working paper},
  langid = {english}
}

@unpublished{naritaMatchMismatchLearning2018,
  title = {Match or Mismatch? Learning and Inertia in School Choice},
  shorttitle = {Match or Mismatch?},
  author = {Narita, Yusuke},
  year = {2018},
  month = jun,
  note = {Working paper},
  number = {3198417},
  address = {Rochester, NY},
  doi = {10.2139/ssrn.3198417},
  urldate = {2023-07-10},
  langid = {english}
}

@unpublished{sweatEndogenousPriorityCentralized,
  title = {Endogenous Priority in Centralized Matching Markets: The Design of the Heart Transplant Waitlist},
  author = {Kurt R. Sweat},
  year = {2024},
  note = {Working paper}
}

@article{verdierWelfareEffectsDynamic2022,
  title = {Welfare {{Effects}} of {{Dynamic Matching}}: {{An Empirical Analysis}}},
  shorttitle = {Welfare {{Effects}} of {{Dynamic Matching}}},
  author = {Verdier, Valentin and Reeling, Carson},
  year = {2022},
  month = mar,
  journal = {The Review of Economic Studies},
  volume = {89},
  number = {2},
  pages = {1008--1037},
  issn = {0034-6527},
  doi = {10.1093/restud/rdab048},
  urldate = {2024-01-09}
}

@article{waldingerTargetingInKindTransfers2021a,
  title = {Targeting {{In-Kind Transfers}} through {{Market Design}}: {{A Revealed Preference Analysis}} of {{Public Housing Allocation}}},
  shorttitle = {Targeting {{In-Kind Transfers}} through {{Market Design}}},
  author = {Waldinger, Daniel},
  year = {2021},
  month = aug,
  journal = {American Economic Review},
  volume = {111},
  number = {8},
  pages = {2660--2696},
  issn = {0002-8282},
  doi = {10.1257/aer.20190516},
  urldate = {2023-12-22},
  langid = {english}
}

@article{yamaguchiEffectsParentalLeave2019,
  title = {Effects of Parental Leave Policies on Female Career and Fertility Choices},
  author = {Yamaguchi, Shintaro},
  year = {2019},
  journal = {Quantitative Economics},
  volume = {10},
  number = {3},
  pages = {1195--1232},
  issn = {1759-7331},
  doi = {10.3982/QE965},
  urldate = {2023-07-16},
  copyright = {Copyright {\copyright} 2019 The Author.},
  langid = {english}
}

@article{yamaguchiEffectsSubsidizedChildcare2018,
  title = {Effects of Subsidized Childcare on Mothers' Labor Supply under a Rationing Mechanism},
  author = {Yamaguchi, Shintaro and Asai, Yukiko and Kambayashi, Ryo},
  year = {2018},
  month = dec,
  journal = {Labour Economics},
  volume = {55},
  pages = {1--17},
  issn = {0927-5371},
  doi = {10.1016/j.labeco.2018.09.002},
  urldate = {2023-07-16},
  langid = {english}
}

@article{yamaguchiHowDoesEarly2018,
  title = {How Does Early Childcare Enrollment Affect Children, Parents, and Their Interactions?},
  author = {Yamaguchi, Shintaro and Asai, Yukiko and Kambayashi, Ryo},
  year = {2018},
  month = dec,
  journal = {Labour Economics},
  volume = {55},
  pages = {56--71},
  issn = {0927-5371},
  doi = {10.1016/j.labeco.2018.08.006},
  urldate = {2023-07-16},
  langid = {english}
}

@article{fackTruthTellingPreferenceEstimation2019,
  title = {Beyond {{Truth-Telling}}: {{Preference Estimation}} with {{Centralized School Choice}} and {{College Admissions}}},
  shorttitle = {Beyond {{Truth-Telling}}},
  author = {Fack, Gabrielle and Grenet, Julien and He, Yinghua},
  year = {2019},
  month = apr,
  journal = {American Economic Review},
  volume = {109},
  number = {4},
  pages = {1486--1529},
  issn = {0002-8282},
  doi = {10.1257/aer.20151422},
  urldate = {2024-01-24},
  langid = {english}
}

@unpublished{leeDynamicAllocationPublica,
  title = {The Dynamic Allocation of Public Housing: Policy and Spillovers},
  author = {Lee, Kwok Hao and Ferdowsian, Andrew and Yap, Luther},
  year = {2024},
  note = {Working Paper},
  langid = {english}
}

\clearpage

\end{document}